\documentstyle[eqsecnum,aps,epsf,multicol]{revtex}

\begin{document}
\draft
\preprint{Submitted to Eur.\ Phys.\ J.\ E}


\title{Adhesion-induced lateral phase separation in membranes}

\author{
Shigeyuki Komura$^{1,2}$\cite{komura}
and David Andelman$^2$
}

\address{
$^1$Department of Materials and Interfaces,
Weizmann Institute of Science,
Rehovot 76100, Israel
}

\address{
$^2$School of Physics and Astronomy,
Raymond and Beverly Sackler Faculty of Exact Sciences, \\
Tel Aviv University, Ramat Aviv 69978, Tel Aviv, Israel
}

\date{\today}
\maketitle


\begin{abstract}
Adhesion between membranes is studied using a phenomenological model,
where the inter-membrane distance is coupled to the
concentration of sticker molecules on the membranes.
The model applies to both adhesion of two flexible membranes and
to adhesion of one flexible membrane onto a second membrane
supported on a solid substrate.
We mainly consider the case where the sticker molecules form bridges
and adhere directly to both membranes.
The calculated mean-field phase diagrams show an upward shift of the
transition temperature indicating that the lateral phase separation
in the membrane is {\it enhanced} due to the coupling effect.
Hence the possibility of adhesion-induced lateral phase separation
is predicted.
For a particular choice of the parameters, the model exhibits a
tricritical behavior.
We also discuss the non-monotonous shape of the inter-membrane
distance occurring when the lateral phase separation takes place.
The inter-membrane distance relaxes to the bulk values with two
symmetric overshoots.
Adhesion mediated by other types of stickers is also considered.
\end{abstract}

\section{Introduction}
\label{sec:intro}

Adhesion of membranes and vesicles has attracted considerable
experimental and theoretical interest because of its prime importance
to many bio-cellular processes \cite{Alberts,Sackmann}.
Theoretical treatments of membranes composed of single component
lipid bilayers have revealed that generic interactions such as van der
Waals, electrostatic or hydration interactions govern the adhesive
properties of interacting membranes.
It is also worthwhile to mention that
related phenomena are found in  unbinding transition of nearly
flat membranes \cite{LL} or adhesion of vesicles to surfaces \cite{SL}.

In addition to general non-specific interactions
mentioned above, it is known from the works of Bell and coworkers
\cite{Bell,BDB} as well as others \cite{Evans},
that highly
specific molecular interactions play an essential role in
biological adhesion.
This interaction acts between complementary pairs of proteins
such as ligand and receptor, or antibody and antigen.
Well studied example of such coupled systems is the biotin-avidin complex.
The avidin molecule has four biotin binding sites, two on each side,
and forms a five-molecules biotin-avidin-biotin complex.
The resulting specific interaction is highly
local and short-ranged.
Measurements by surface force apparatus \cite{LISK} or atomic force
microscopy \cite{FMG,MFG} have shown that the  force required to
break a biotin-avidin bond is about 170pN.
In related experiments measuring chemical equilibrium constants \cite{M},
it was found that
the biotin-avidin binding energy is about 30--35 $k_BT$ which is larger
than thermal fluctuations. Other coupled systems are those of selectins and
their sugar ligands where the bond is much weaker, of the order of
5$k_B T$ \cite{ZB95,ZB98}.

More recently several models taking into account thermal fluctuations in
membrane adhesion have been proposed. Zuckerman and Bruinsma
\cite{ZB95,ZB98} used a statistical mechanics model which is  mapped
onto a two-dimensional Coulomb plasma with attractive interactions.
They predicted an enhancement of the membrane adhesion due to thermal
fluctuations.
In another work, Lipowsky considered the adhesion of lipid
membranes which includes anchored stickers, {\it i.e.}, anchored molecules
with adhesive segments \cite{Lipowsky96,Lipowsky97}. It was shown that
flexible membranes can adhere if the sticker concentration exceeds a
certain threshold.
If the multi-component membranes, including lipids and
sticker molecules, undergo a phase separation, the adhesion is dominated
by the sticker-rich domains.
Further studies in this direction using
mean-field theory and Monte Carlo simulations \cite{WNL} obtained a phase
separation which is driven both by attractive intra-membrane sticker
interactions and fluctuation-induced interactions between stickers.

The problem of multi-component membrane adhesion is intimately related to
that of formation of domains (a lateral phase separation).
This has been observed by several experiments.
For example, the biotin-avidin interaction occurring during vesicle-vesicle
adhesion was investigated by a micropipette technique \cite{NN}.
The adhesion between one avidin-coated vesicle and a second biotinylated
vesicle is followed by an accumulation of biotin-avidin complex in the
contact zone.
This accumulation of cross-bridges between the two vesicles
is found to be a diffusion-controlled
process.

Adhesion-induced phase separation has been observed
by Albersd\"orfer {\it et al.} and results from the interplay
between long-range repulsive and short-range attractive interactions
\cite{AFS,ABS}.
The membrane includes repeller molecules in the form of
lipopolymers (modified DOPE lipid with a polyethyleneoxide headgroup),
mimicking glycocalix in real biological systems. The other component
is a receptor molecule in the form of biotinylated lipids (DOPE-X-biotin).
This  lipopolymer is responsible for longer-range repulsive interaction,
while the short-range attractive interaction is introduced
by adding streptavidin to the extra-cellular solution. The streptavidin
acts as a connector between the biotinylated lipids on the two membranes.
A technique of reflection interference contrast microscopy \cite{RFSS}
was used to observe domain formation on a vesicle adhering to a membrane
supported on a solid substrate.
The lateral phase separation
on both membranes leads to the formation of domains of tight adhesion
separated by domains of loose adhesion \cite{AFS,ABS}.

In a related work, adhesion between cationic vesicles and
anionic supported membranes revealed that electrostatic interactions
induce lateral charge segregation on the membrane \cite{NFBS,NBS}.
This phase separation leads to patches of tight inter-membrane
contact and decoupled
``blisters''.
Furthermore, adhesion of membranes including self-recognizing
homophilic molecules and lipopolymers has been investigated
\cite{BBS}.
It was found that the initial weak adhesion is followed by slower
aggregation into tightly bound domains coexisting  with
domains of weak adhesion.
The result has been interpreted in terms of a double-well
inter-membrane interaction potential due to the presence of
the lipopolymers.
Let us emphasize that
in all the above mentioned experiments, it was
reported that adhesion molecules aggregate
spontaneously and form domains of tight adhesion.

It is generally believed that multi-component biomembranes
in physiological conditions are close to their
critical point, and membrane functions are partially governed through
phase separation processes.
Moreover, concentration fluctuations in the vicinity of the critical
point may affect biophysical properties of membranes and
can be of importance in regulating membrane processes in a robust way.
Recently this conjecture was supported by an experiment of an
insoluble Langmuir monolayer at the air/water interface \cite{KPHM,KM}.
The monolayer was prepared in two different steps.
The first mimics the composition of the inner leaflet of a cell
biomembrane, while the second mimics the outer leaflet.
In both cases, by using fluorescence microscopy technique, it was found
that the Langmuir monolayer is close to its corresponding critical point
of demixing.

So far, the interplay between lateral phase separation and
membrane adhesion has not been considered theoretically in detail
except in Refs.\ \cite{Lipowsky96,Lipowsky97,WNL}.
The work in Refs.\ \cite{NFBS,NBS} deal only with the specific
case of oppositely charged membranes.
In this paper we provide a general phenomenological approach for
the adhesion of multicomponent membranes.
Using a mean-field theory, we investigate how the lateral phase
separation within the membrane is affected by the adhesion of
membranes.
Like in Refs.\ \cite{Lipowsky96,Lipowsky97,WNL}, we consider adhesion
mediated by sticker molecules.
Sticker molecules are polymers or macromolecules anchored to one
membrane and interacting with the other membrane by another sticky
part of the molecule.
They can form bridges between two adjacent membranes
(so called {\it trans}-interaction) \cite{Lipowsky96},
and play an essential role in the adhesion of cell membranes
in biological systems.

We distinguish three types of adhesion depending on the structure
of bridges as represented in Fig.\ \ref{figtype}. (i)
``Bolaform-sticker" adhesion where each bridge molecule consists
of a single sticker having two sticky ends (Fig.\
\ref{figtype}(a)). One sticker end is anchored to one membrane
while the other end is adhering directly to the second membrane.
(ii) ``Homophilic-sticker" adhesion where the bridges are formed
by two stickers of the same type (Fig.\ \ref{figtype}(b)). Each
sticker is anchored on one of the membranes, while their free ends
bind together to form the bridge. (iii) ``Lock-and-key" adhesion
where the bridges consist of two different stickers forming a
ligand-receptor type bond (Fig.\ \ref{figtype}(c)). This case
represents an asymmetric adhesion due to the lack of symmetry
between the ligand and receptor. In the present work, we mainly
discuss the symmetric bolaform-sticker adhesion (case (i) above)
using a model where the equilibrium spacing between two membranes
is coupled to the local concentration of stickers. Even in the
latter symmetric case, a certain asymmetry can be obtained by
controlling separately the sticker chemical potentials on the two
membranes.
An important consequence of our model is that the lateral phase
separation is {\it enhanced}.

This paper is organized as follows.
In the next section, we explain our phenomenological model
of bolaform-sticker adhesion.
The mean-field phase diagrams are given in Sec.\ \ref{sec:phase}.
The inter-membrane distance  between two coexisting domains
is calculated in Sec.\ \ref{sec:profile}.
Finally discussion is provided in Sec.\ \ref{sec:discussion}
where the other types of adhesion mentioned above are considered.

\section{Bolaform-sticker adhesion}
\label{sec:single}

In this section, we treat the case where the adhesion is mediated
by a single type of sticker molecules which are anchored irreversibly
to one membrane and stick to the other membrane by another sticky 
part of the molecules as in Fig.\ \ref{figtype}(a).
The anchor segments consist of a hydrophobic segment and 
penetrate into the hydrophobic interior of the lipid bilayer.
The sticky segments, on the other hand, adhere directly to another 
membrane having some potential of sticking \cite{Lipowsky97}.  
As mentioned above, we call this a ``bolaform-sticker''. Consider
two interacting membranes labeled by $i=1,2$ consisting of lipid
molecules and bolaform-stickers as schematically shown in Fig.\
\ref{figmem}. Let the sticker concentration in each membrane be
denoted by $\psi_i(\mbox{\boldmath $r$})$, where
$\mbox{\boldmath $r$}=(x,y)$ is a two-dimensional planar vector
and $0 \le \psi_i(\mbox{\boldmath $r$}) \le 1$. Note that the
average concentrations of stickers on the two membranes, $\langle
\psi_1\rangle$ and $\langle \psi_2\rangle$, do not have to be the
same.

When the adhesion molecules are very flexible, they can bend
back to form arches on a single membrane.
In order to avoid such a situation in experiments and in the model, the
bending rigidity of the sticker molecules should be sufficiently large.
Hence we assume that for stiff enough stickers all the bonds are
inter-membrane ones connecting the two separate membranes
as considered in Refs.\ \cite{Lipowsky96,Lipowsky97}.

The interaction  between two stickers on the same membrane is
called {\it cis}-interaction and can be repulsive or attractive.
Here we discuss the case in which this interaction is attractive.
Then, below a certain critical temperature, the multi-component
membrane undergoes a first-order phase transition and stickers
form lateral domains. As shown in Fig.\ \ref{figphase}, a
sticker-poor phase coexists with a sticker-rich phase in the
two-phase region of the phase diagram. The sticker critical
concentration $\psi_c$ and the critical temperature $T_c$ are
assumed to be the same for the two planar membranes. We define the
concentration difference $\phi_i(\mbox{\boldmath $r$})$ for each
of the membranes with respect to the critical concentration by
\begin{equation}
\phi_i(\mbox{\boldmath $r$}) = \psi_i(\mbox{\boldmath $r$}) -
\psi_c~~~~~(i=1,2), \label{difference}
\end{equation}
where $\phi_i$  varies between $-1 \le \phi_i \le 1$.

The total free energy of the two coupled membranes can be written as
a sum of several terms detailed below.
The first contribution describes the lateral phase separation
of each membrane.
Motivated by recent experiments on Langmuir monolayers
\cite{KPHM,KM} demonstrating that the inner and outer leaflets
of biomembranes are close to their critical point, we employ a
phenomenological Ginzburg-Landau free energy which is an expansion in
powers of the order parameters $\{\phi_i\}$.
Hence we have
\begin{equation}
F_1= \frac{1}{2} \sum_{i=1,2} \int \mbox{d}^2 \mbox{\boldmath
$r$} \, \left[ \frac{1}{2} c (\nabla \phi_i)^2 + \frac{1}{2}t
\phi_i^2 + \frac{1}{4} \phi_i^4 -\mu_i \phi_i \right].
\label{glene}
\end{equation}
This expansion for the free energy can be justified close to a
critical point where the $\phi_i$'s are small enough. The
parameter $c$ representing the line tension acting at the domain
boundary, and the reduced temperature $t = (T-T_c)/T_c$ are taken
to be the same for the two membranes. On the other hand, the
chemical potential $\mu_i$, coupled to the membrane sticker
concentration $\phi_i$, can differ between the two membranes since
the sticker concentrations on the two membranes do not have to be
the same. We recall that each bolaform-sticker is modeled with one
of its ends anchored irreversibly to one membrane, while the
second sticky end is attracted to the second membrane. The sticker
concentration is associated with the anchored end of the stickers.
The normalization factor 1/2 in (\ref{glene})
is introduced in order to write down the
free energy per single membrane.
The coefficient of the fourth order term can generally be
set as a positive constant without loss of generality.

It is convenient to introduce the following new variables
for the average and the difference between the two concentrations:
\begin{equation}
\phi_+ = \frac{\phi_2+\phi_1}{2}=\frac{\psi_2+\psi_1}{2}-\psi_c,~~~~~
\phi_- = \frac{\phi_2-\phi_1}{2}=\frac{\psi_2-\psi_1}{2},
\end{equation}
where $-1 \le \phi_+ \le 1$ and $-1 \le \phi_- \le 1$.
In terms of these new variables, (\ref{glene}) can be written as
\begin{eqnarray}
F_1 = & & \int \mbox{d}^2 \mbox{\boldmath $r$} \, \biggl[
\frac{1}{2} c \left[ (\nabla \phi_+)^2 + (\nabla \phi_-)^2 \right]
+\frac{1}{2} t (\phi_+^2 + \phi_-^2) \nonumber  \\ & & +
\frac{1}{4} (\phi_+^4 + 6 \phi_+^2 \phi_-^2 + \phi_-^4) -\mu_+
\phi_+ - \mu_- \phi_- \biggr], \label{glene2}
\end{eqnarray}
where
\begin{equation}
\mu_+ = \frac{\mu_2+\mu_1}{2},~~~~~
\mu_- = \frac{\mu_2-\mu_1}{2}.
\end{equation}
The chemical potential $\mu_-$ associated with the order parameter
$\phi_-$ is non-zero when the symmetry between the two membranes is
explicitly broken.
Namely, the two interacting membranes have different average
concentrations of stickers.

Next we consider the out-of-plane deformation energy of the two
membranes. As depicted in Fig.\ \ref{figmem}, the  membrane shape
is parameterized by their heights $\ell_1(\mbox{\boldmath $r$})$,
$\ell_2(\mbox{\boldmath $r$}')$, above the  $x$-$y$ reference
plane. Working in the Monge representation it is implicitly
assumed that the membranes remain flat on average and have no
overhangs. This approach can be also useful to treat adhesion of
vesicles in their contact zone. When the vesicle is large enough,
it will be roughly flat close to the contact region, and the
entire vesicle can be thought of as a reservoir for the stickers.
Returning to the deformation energy, it can be written as the sum
of the bending energy and the surface tension of each of the two
membranes separately, as well as the interacting potential energy
between them \cite{Safran,BGP}:
\begin{eqnarray}
F_2 = & &  \frac{1}{2} \sum_{i=1,2} \int \mbox{d}^2
\mbox{\boldmath $r$} \, \left[ \frac{1}{2} \kappa (\nabla^2
\ell_i)^2  + \frac{1}{2} \sigma (\nabla \ell_i)^2  \right]
\nonumber \\ & & + \int \mbox{d}^2 \mbox{\boldmath $r$}~
v(\ell_1-\ell_2; \psi_1, \psi_2), \label{elene1}
\end{eqnarray}
where $\kappa$ is the bending rigidity, $\sigma$ is the mechanical
surface tension acting on the membranes, and $v$ is the potential
energy per unit area representing the inter-membrane interactions.
For simplicity, $\kappa$ and $\sigma$ are assumed to be equal for the two
membranes and do not vary as a function of the sticker concentration
$\psi_i$.
The potential $v(\ell_1-\ell_2; \psi_1, \psi_2)$ can be generally 
assumed to be a function of the local relative height coordinate 
$\ell_1-\ell_2$ and the sticker concentration $\psi_i$.
The former assumption is the so-called Derjaguin approximation \cite{Israel}.
The dependence on the sticker concentration $\psi_i$ will be considered later.

We now make a change of variables and
transform to the center-of-mass and relative coordinates
given, respectively,  by
\begin{equation}
L = \frac{\ell_2+\ell_1}{2},~~~~~\ell = \frac{\ell_2-\ell_1}{2}.
\end{equation}
Only terms which depend on $\ell$ can be considered
in the case where the center-of-mass is stationary, hence
$L$ is a constant of motion.
Then (\ref{elene1}) can be written as \cite{rigidity}
\begin{equation}
F_2 = \int \mbox{d}^2 \mbox{\boldmath $r$} \, \left[ \frac{1}{2}
\kappa (\nabla^2 \ell)^2  + \frac{1}{2} \sigma (\nabla \ell)^2  +
v(\ell; \psi_1, \psi_2) \right]. \label{elene2}
\end{equation}

In the high-temperature phase, the stickers are homogeneously
distributed, and each of the membrane is in a one-phase region on
the phase diagram. We assume that even in the absence of sticker
molecules, such membranes are bound to each other due to the
balance between the short-range repulsive ({\it e.g.}, hydration
interaction) and longer-range attractive interactions ({\it e.g.},
van der Waals interaction). Hence, we do not consider the
interesting problem of the unbinding transition
\cite{Lipowsky96,Lipowsky97,WNL}. Although the membranes are always
bound together, their equilibrium distance $\ell$ depends on the
sticker concentration.
Let us consider the potential $v(\ell; \psi_1, \psi_2)$ in 
(\ref{elene2}) for $\phi_+=0$. Note that $\phi_+=0$ means that
$(\psi_1+\psi_2)/2=\psi_c$, namely, the average sticker
concentration on the two membranes is at its critical value
$\psi_c$. The inter-membrane potential $v(\ell; \phi_+=0)$ is assumed to
have a single minimum at a certain inter-membrane distance
$\ell=\ell_0$ for $\phi_+=0$. This gives the equilibrium distance
between the two bound membranes. The deviation of the
inter-membrane distance from $\ell_0$ is defined by the
dimensionless quantity $\delta(\mbox{\boldmath $r$})$ given by
\begin{equation}
\delta(\mbox{\boldmath $r$}) = \frac{\ell(\mbox{\boldmath $r$})
- \ell_0}{\ell_0}. \label{scaledell}
\end{equation}
For small deviations from the minimum of the potential,
$v(\ell; \phi_+=0)$ can be expanded to second order.
This is known as the harmonic approximation and gives
\begin{eqnarray}
v(\ell; \phi_+=0) & \approx & v(\ell_0) + 
\frac{1}{2}v''(\ell_0) (\ell - \ell_0)^2
\nonumber  \\
& = & v(\ell_0) + \frac{1}{2}V \delta^2,
\label{taylor}
\end{eqnarray}
where $v''(\ell_0)$ is the second derivative of $v$ with respect to $\ell$
evaluated at $\ell = \ell_0$, $V \equiv  v''(\ell_0) \ell_0^2$, and
$v'(\ell)=0$ at $\ell=\ell_0$ \cite{numbers}.
Using (\ref{scaledell}) and (\ref{taylor}), (\ref{elene2}) can be written as
\begin{equation}
F_2 \approx \int \mbox{d}^2 \mbox{\boldmath $r$} \, \left[
\frac{1}{2} K (\nabla^2 \delta)^2  + \frac{1}{2} \Sigma (\nabla
\delta)^2  + \frac{1}{2}V \delta^2 \right], \label{elene3}
\end{equation}
with $K \equiv \kappa \ell_0^2$ and $\Sigma \equiv\sigma \ell_0^2$.
This is the expression of the deformation energy within the harmonic
approximation and it served as a starting point to many calculations
on membrane adhesion \cite{SL,BGP,SA}.

Now we will include the effect of the adhesion on the phase
separation and suggest a lowest-order coupling between the
composition $\phi_i(\mbox{\boldmath $r$})$ and the inter-membrane
distance $\delta(\mbox{\boldmath $r$})$. When the membranes are
quenched into a two-phase region of the phase diagram a
sticker-poor phase coexists with a sticker-rich phase. As shown in
Fig.\ \ref{figsym}, this can lead to different inter-membrane
distance for the different membrane domains. Since the sticky
segments of the bridges adhere directly onto the two membranes,
the coupling is proportional to the sum of the local sticker
concentrations of the two membranes. This can be
phenomenologically represented by the following coupling term:
\begin{equation}
F_3 = \frac{\alpha}{2\ell_0} \int \mbox{d}^2 \mbox{\boldmath $r$}
\, (\psi_1+\psi_2) \ell =\alpha \int \mbox{d}^2 \mbox{\boldmath
$r$} \, \phi_+ \delta \, + \, \cdots, \label{symcoup}
\end{equation}
where the coupling constant $\alpha$ is positive preferring smaller
separation $\delta<0$ in regions where the average concentration
$\phi_+$ is positive (or $\psi_1+\psi_2 >2 \psi_c$).
In the last expression of (\ref{symcoup}), we have neglected the linear
terms in $\phi_+$ and
$\delta$, which merely shift the chemical potential or minimum of the
potential, respectively.
Depending on the value of $\phi_+$, this coupling term not only
introduces a shift of the minimum of the potential $v$ but
also changes the minimum value of the potential.
Note also that (\ref{symcoup}) is symmetric with respect to the
exchange of the two membranes $1 \leftrightarrow 2$.

The above linear coupling energy can also be understood in the
following way. Let us first consider a single flexible membrane
with sticker molecules adhering to a flat substrate. Suppose
$v_r(\ell)$ and $v_f(\ell)$ are the potentials for sticker-rich
and sticker-free membrane, respectively. Following the same
discussion as in (\ref{taylor}), each of the potential is
parabolic around a different separation: $v_r \approx
a(\ell-\ell_r)^2$ and $v_f \approx a(\ell-\ell_f)^2$. The
effective potential can be obtained by a linear combination of
these two potentials, {\it i.e.}, $(1-\psi) v_f(\ell)+\psi
v_r(\ell)$. By expanding $v_f(\ell)$ and $v_r(\ell)$, we get a
coupling term which is proportional to $\psi \ell$. In the case of
adhesion between two membranes, we add the contributions from both
of the membranes and obtain the coupling energy as given in
(\ref{symcoup}). The same argument can be repeated for any
arbitrary adhesion potentials, $v_f(\ell)$ and $v_r(\ell)$,
provided each of them has a single well-defined minimum at some
distance $\ell$.

The total free energy considered in our model is the sum of
(\ref{glene2}), (\ref{elene3}), and (\ref{symcoup}):
\begin{equation}
F = F_1 + F_2 + F_3.
\label{symtotal}
\end{equation}
Here it is convenient to convert to Fourier space. The Fourier
transform of any function $f(\mbox{\boldmath $r$})$ is defined as
\begin{equation}
\tilde{f}(\mbox{\boldmath $q$}) = \int \mbox{d}^2
\mbox{\boldmath $r$} \, f(\mbox{\boldmath $r$}) e^{i
\mbox{\boldmath $q$} \cdot \mbox{\boldmath $r$}},
\end{equation}
where $\mbox{\boldmath $q$}$ is the two-dimensional in-plane
wavevector. The total free energy can be expressed as
\begin{eqnarray}
F & = & \frac{1}{A}\sum_{\mbox{\boldmath $q$}} \, \biggl[
\frac{1}{2} (t +c q^2) \big( \vert \tilde{\phi}_+(\mbox{\boldmath
$q$}) \vert^2 + \vert \tilde{\phi}_-(\mbox{\boldmath $q$})
\vert^2 \big) \nonumber  \\ & & +  \frac{1}{2} (V + \Sigma q^2 + K
q^4) \vert \tilde{\delta}(\mbox{\boldmath $q$}) \vert^2 + \alpha
\tilde{\phi}_+(\mbox{\boldmath $q$})
\tilde{\delta}(-\mbox{\boldmath $q$}) \biggr] \nonumber  \\ & & +
\int \mbox{d}^2 \mbox{\boldmath $r$} \, \biggr[ \frac{1}{4}
(\phi_+^4 + 6 \phi_+^2 \phi_-^2 + \phi_-^4) \nonumber  \\ & &
-\mu_+ \phi_+ -\mu_- \phi_- \biggr], \label{nonlocal}
\end{eqnarray}
where $A$ is the area of the membranes projected on the $x$-$y$ plane.
For convenience the free energy (\ref{nonlocal}) is written as
a combination of real space and Fourier space terms.

Within the mean-field level, the free energy in terms of $\phi_+$
and $\phi_-$ is obtained by functionally minimizing $F$ with
respect to $\tilde{\delta}(\mbox{\boldmath $q$})$. Then we find
\begin{equation}
\tilde{\delta}(\mbox{\boldmath $q$}) = -\frac{\alpha
\tilde{\phi}_+(\mbox{\boldmath $q$})} {V + \Sigma q^2 + K q^4}.
\label{deltaq}
\end{equation}
Hence the inter-membrane distance $\delta=(\ell-\ell_0)/\ell_0$ is
fully determined by the value of $\phi_+$.
By inserting (\ref{deltaq}) into (\ref{nonlocal}), the resulting
free energy depends only on $\phi_+$ and $\phi_-$, and becomes
\begin{eqnarray}
F & = & \frac{1}{A}\sum_{\mbox{\boldmath $q$}} \, \biggr[
\frac{1}{2} \tilde{\Gamma}_+(\mbox{\boldmath $q$}) \vert
\tilde{\phi}_+(\mbox{\boldmath $q$}) \vert^2 + \frac{1}{2} (t + c
q^2) \vert \tilde{\phi}_-(\mbox{\boldmath $q$}) \vert^2 \biggr]
\nonumber  \\ & & + \int \mbox{d}^2 \mbox{\boldmath $r$} \,
\biggr[ \frac{1}{4} (\phi_+^4 + 6 \phi_+^2 \phi_-^2 + \phi_-^4)
\nonumber  \\ & & -\mu_+ \phi_+ -\mu_- \phi_- \biggr],
\end{eqnarray}
where
\begin{equation}
\tilde{\Gamma}_+(\mbox{\boldmath $q$}) = t + cq^2 -
\frac{\alpha^2}{{V + \Sigma q^2 + K q^4}}. \label{gammap}
\end{equation}
If we expand the last term in (\ref{gammap}) for small $q$,
we obtain
\begin{equation}
\tilde{\Gamma}_+(\mbox{\boldmath $q$}) \approx (t - \gamma) +
\left( c + \frac{\alpha^2 \Sigma}{V^2} \right) q^2,
\label{gammaplus}
\end{equation}
with
\begin{equation}
\gamma\equiv \frac{\alpha^2}{V}.
\label{gamma}
\end{equation}
The parameter $\gamma$ is an important parameter
characterizing the coupling strength.
The first two terms in (\ref{gammaplus}) implies an upward shift of
the transition temperature, as will be discussed in detail in the
next section.
We also find that the presence of the coupling ($\alpha\ne 0$) increases
the line tension $c$ provided the mechanical surface tension
$\Sigma$ is non-zero; $c \rightarrow c + {\alpha^2\Sigma}/{V^2}$.

\section{Phase Diagrams}
\label{sec:phase}

In this Section, we calculate the mean-field phase diagrams for
bolaform-sticker adhesion using the free energy explained in the
previous section. In order to study the bulk properties of the
system, we set $\mbox{\boldmath $q$}=0$ and study the homogeneous
solutions, $\phi_i$'s and $\delta$ being constants. From
(\ref{deltaq}), the inter-membrane distance which minimizes the
free energy is given by
\begin{equation}
\delta = - \frac{\alpha \phi_+}{V}.
\label{deltasym}
\end{equation}
Since $\alpha$ is positive, $\delta$ is negative (smaller
inter-membrane distance) for positive $\phi_+$, and $\delta$ is
positive (larger inter-membrane distance) for negative $\phi_+$.
By substituting back this $\delta$ into the free energy
$f$ per unit area for homogeneous (constant) $\phi_+$ and $\phi_-$,
we obtain
\begin{eqnarray}
f = & & \frac{1}{2} \left(t - \gamma \right) \phi_+^2
+ \frac{1}{2} t \phi_-^2
\nonumber  \\
& & + \frac{1}{4} (\phi_+^4 + 6 \phi_+^2 \phi_-^2 + \phi_-^4)
-\mu_+ \phi_+ -\mu_- \phi_-,
\label{fslocal2}
\end{eqnarray}
where $\gamma$ is defined in (\ref{gamma}).
Notice that $\gamma$ is never negative and vanishes only when $\alpha=0$.
Therefore, when $\mu_+=\mu_-=0$, the field $\phi_+$ will order before
$\phi_-$, and the phase with $\phi_+ \neq 0$ and $\phi_- = 0$ is
expected \cite{CL}.
Although the phase behavior of this free energy
can be examined in general, we concentrate here only on two particular
cuts
in the parameter space, {\it i.e.}, $\mu_- = 0$ and $\mu_+ = 0$.
In these cases one can clearly see the effect of adhesion on the lateral
phase separation.

\subsection{The case $\mu_-=0$}

When $\mu_- = 0$ the
two membranes have the same chemical potential $\mu_1=\mu_2$.
Since the chemical potential $\mu_+$ is coupled to $\phi_+$,
 $f$ can be minimized first with respect to $\phi_-$.
A ``symmetric phase'' is obtained for $t+3\phi_+^2>0$ with
\begin{equation}
\phi_-=0,
\end{equation}
where the two membranes have the same concentrations,
$\phi_1=\phi_2$.
Likewise, two ``asymmetric phases'' are obtained for $t+3\phi_+^2<0$
with
\begin{equation}
\phi_- = \pm \sqrt{-t-3\phi_+^2}.
\end{equation}
In the asymmetric phase the two membranes have  different concentrations
$\phi_1 \neq \phi_2$.
After inserting these expressions into (\ref{fslocal2}) with
$\mu_-=0$, the free energy becomes
\begin{equation}
f_1 = \left\{
\begin{array}{lll}
\frac{1}{2} \left(t - \gamma \right) \phi_+^2
+ \frac{1}{4} \phi_+^4 -\mu_+ \phi_+ & ~~~~~ & \mbox{for~~~~~ $t+3\phi_+^2>0$} \\
- \frac{1}{4}t^2 - \frac{1}{2}\left(2t + \gamma \right) \phi_+^2
- 2 \phi_+^4 -\mu_+ \phi_+ & ~~~~~ & \mbox{for~~~~~ $t+3\phi_+^2<0$}
\end{array} \right..
\label{fsmin2}
\end{equation}
Notice that this free energy is continuous at $t=-3\phi_+^2$.
This free energy $f_1$ can now be minimized with respect to $\phi_+$.
The resulting
equation of state is written as
\begin{equation}
\mu_+ = \left\{
\begin{array}{lll}
\left(t - \gamma \right) \phi_+ + \phi_+^3
& ~~~~~ & \mbox{for~~~~~ $t+3\phi_+^2>0$} \\
- \left(2t + \gamma \right) \phi_+ - 8 \phi_+^3
& ~~~~~ & \mbox{for~~~~~ $t+3\phi_+^2<0$}
\end{array} \right..
\end{equation}
The phase diagram can now be calculated
and the two-phase region is obtained by the Maxwell
construction.
The phase diagram for $\mu_-=0$ is illustrated in Fig.\
\ref{figphasesym}.

For $t > 0$, only the symmetric phase with $\phi_-=0$ can appear
since $t+3\phi_+^2>0$.
Two symmetric phases with different $\phi_+$ can coexist
when $t < \gamma$.
The coexistence curve is simply given by
\begin{equation}
\phi_+ = \pm \sqrt{- t+ \gamma},
\label{coexist}
\end{equation}
and the associated critical point is located at
\begin{equation}
(t, \phi_+, \mu_+)_c =(\gamma, 0, 0).
\label{critical}
\end{equation}
We stress that
the critical temperature is {\it increased}
from $t_c=0$ to $t_c=\gamma = \alpha^2/V$ due to the coupling between the
composition $\phi_+$ and the inter-membrane distance $\delta$ as given
in (\ref{symcoup}).
In other words, the phase separation is {\it enhanced} by the adhesion
of membranes.
As presented in Fig.\ \ref{figsym}, the two coexisting
values of $\phi_+$ given by (\ref{coexist}) lead to different
inter-membrane distances $\delta$ according to (\ref{deltasym}).
Since $\alpha >0$,
$\delta$ is negative ($\ell < \ell_0$) in the sticker-rich domain,
and this phase is called the ``tight phase'' (T).
On the other hand, $\delta$ is positive ($\ell > \ell_0$)
in the sticker-poor domain and this phase is called the
``loose phase'' (L).
However, for each of the coexisting tight and loose phases, $\phi_-=0$,
which means that the sticker concentration is the same in the two
membranes, $\phi_1=\phi_2$.

For $-\gamma/2 < t < 0$, the asymmetric phase with $\phi_- \neq 0$
is always unstable, and the tight and loose phases coexist according to
(\ref{deltasym}) and (\ref{coexist}).
For $t < -\gamma/2$, the asymmetric phase can be locally stable but it is
only metastable.
Namely, its free energy is higher than that of the symmetric phase.
Hence the coexistence between the tight and loose phases
given by (\ref{deltasym}) and (\ref{coexist}) preempts the asymmetric phase.
The limit of metastability of the asymmetric phase is obtained by calculating
the second derivative of the second equation of (\ref{fsmin2}) with respect
to $\phi_+$.
This leads to
\begin{equation}
\phi_+ = \pm \sqrt{\frac{- 2t- \gamma}{24}},
\label{meta}
\end{equation}
which is also shown as a dotted line inside the L+T coexisting
region of Fig.\ \ref{figphasesym}(a).

In summary, for $\mu_-=0$, the asymmetric phase $\phi_- \neq 0$
does not exist as a stable phase for any temperature.
At most it is metastable and occurs within the L+T coexistence region.
The tight and loose phases coexist for $t < \gamma =\alpha^2/V$
according to (\ref{deltasym}) and (\ref{coexist}).

\subsection{The case $\mu_+=0$}

Next we consider the case of $\mu_+=0$ but with
$\mu_- \neq 0$.
This means that the chemical potentials of the two membranes
have the same magnitude but opposite sign, {\it i.e.},
$\mu_1 = -\mu_2$.
This is a special case of the more general situation where
the symmetry between the two membranes is explicitly broken.
Now $f$ in (\ref{fslocal2}) can be minimized with respect to
$\phi_+$ first.
As long as $t+3\phi_-^2>\gamma$, the only solution is
\begin{equation}
\phi_+=0.
\end{equation}
This is called the ``middle phase'' (M) where the inter-membrane
distance is exactly $\ell_0$ (or $\delta=0$).
Again note that $\phi_+=0$ means that $\psi_1+\psi_2=2 \psi_c$.
For $t+3\phi_-^2<\gamma$, we have the tight (or loose) phase with
\begin{equation}
\phi_+ = \pm
\sqrt{-t + \gamma -3\phi_-^2},
\label{asymcoexist}
\end{equation}
where $\ell$ deviates from $\ell_0$ (or $\delta \neq 0$) according to
(\ref{deltasym}).
Since $\mu_+=0$, both the tight and the loose phases are energetically
degenerated and they coexist.
By substituting $\phi_+$ back into (\ref{fslocal2}) with $\mu_+=0$,
the free energy becomes
\begin{equation}
f_2 = \left\{
\begin{array}{lll}
\frac{1}{2}t \phi_-^2 + \frac{1}{4} \phi_-^4 - \mu_- \phi_-
& ~~~~~ & \mbox{for~~~~~ $t+3\phi_-^2>\gamma$} \\
-\frac{1}{4} \left(t - \gamma \right)^2 +
\frac{1}{2} \left( -2 t+ 3\gamma \right) \phi_-^2
- 2 \phi_-^4  - \mu_- \phi_-
& ~~~~~ & \mbox{for~~~~~ $t+3\phi_-^2<\gamma$}
\end{array} \right..
\end{equation}
After minimizing with respect to $\phi_-$, the equation of state is
given as
\begin{equation}
\mu_- = \left\{
\begin{array}{lll}
t \phi_- + \phi_-^3
& ~~~~~ & \mbox{for~~~~~ $t+3\phi_-^2>\gamma$} \\
\left( -2t + 3 \gamma \right) \phi_- - 8 \phi_-^3
& ~~~~~ & \mbox{for~~~~~ $t+3\phi_-^2<\gamma$}
\end{array} \right..
\end{equation}

The calculated phase diagrams for $\mu_+=0$ are shown
in Fig.\ \ref{figphaseasym}.
The phase diagram is symmetric about $\phi_-=0$ and
$\mu_- = 0$ as a consequence of the $\phi_+^2 \phi_-^2$
coupling term, and lack of any odd terms in $\phi_+$ in the free energy.
For $t > \gamma$ there is a one-phase region of the
middle phase with $\phi_+=0$ since $t+3\phi_-^2>\gamma$.
For $5\gamma/6 < t < \gamma$, the system undergoes a second-order phase
transition between the middle phase ($\phi_+=0$) and the tight (or loose)
phase ($\phi_+ \neq 0$).
The analytical expressions of the second-order phase transition lines are
\begin{equation}
\phi_- = \pm \sqrt{\frac{-t+\gamma}{3}},
\end{equation}
and
\begin{equation}
\mu_- = \pm \frac{2t+\gamma}{3}\sqrt{\frac{-t+\gamma}{3}},
\end{equation}
in Fig.\ \ref{figphaseasym}, respectively.
On the second-order phase transition line, $\phi_+$ goes continuously
to zero.

For $t < 5\gamma/6$, the transition changes to first order.
This has been numerically determined by the Maxwell construction.
The point which connects the first- and second-order phase transition
lines is a tricritical point \cite{CL}.
In our model, it is located at
\begin{equation}
(t,\phi_-, \mu_-)_{tcp}=\left( \frac{5}{6}\gamma,~
\pm \frac{1}{3\sqrt{2}}\gamma^{1/2},~
\pm \frac{4\sqrt{2}}{27}\gamma^{3/2} \right).
\end{equation}
The first-order phase transition corresponds to the coexistence of
the middle phase with $\phi_+=0$ and the tight (or loose) phase
with $\phi_+ \neq 0$. The obtained two-phase coexistence region is
indicated by ``M+T'' in Fig.\ \ref{figphaseasym}(a). Within the
present Ginzburg-Landau expansion, the tight phase persists even
if we go to low temperatures. Because of the degeneracy between
tight and loose phases, the first-order line near the tricritical
point actually corresponds to coexistence of three phases: tight,
loose, and middle phases.

In continuation to the discussion of the previous subsection
($\mu_-=0$), we see that the phase separation is also {\it
enhanced} for the $\mu_+=0$ parameter space. It occurs at higher
temperatures, since the tricritical temperature $t_{tcp}=
5\gamma/6 = 5\alpha^2/6V$ is positive for $\alpha \neq 0$. It is
important to notice that in the middle phase with $\phi_+=0$, the
inter-membrane distance is $\ell_0$ since $\delta=0$. On the other
hand, in the tight (loose) phase with $\phi_+ > 0$ ($\phi_+ < 0$),
according to (\ref{deltasym}) and (\ref{coexist}), $\ell < \ell_0$
($\ell > \ell_0$). The coexisting membrane domains between tight
and middle phases, or between loose and middle phases is
schematically represented in Fig.\ \ref{figasym}.

We end this section by commenting on the general case when both
$\mu_+$ and $\mu_-$ are non-zero. When $\mu_+$ becomes non-zero,
the degeneracy between the tight and the loose phases is lifted.
In such a case, instead of the three-phase coexistence for
$\mu_+=0$, there is a coexistence between either tight and middle
phases, or between loose and middle phases as shown in Fig.\
\ref{figasym}(a) and (b), respectively. Notice that the
tricritical point exists only when $\mu_+=0$. In a more general
phase diagram drawn in the ($t,\mu_+,\mu_-$) space, three
second-order lines meet at the tricritical point. In the
three-dimensional parameter space, these second-order lines lie on
the perimeter of two-phase coexistence planes between either tight
and middle phases (T+M) or between loose and middle phases (L+M).

\section{Non-Monotonous Membrane Profile}
\label{sec:profile}

One of our assumptions was
that the inter-membranes potential $v(\ell; \phi_+=0)$
has a single minimum at $\ell = \ell_0$ when $\phi_+=0$.
In the absence of thermal fluctuations, two homogeneous membranes are
bound with inter-membrane distance $\ell_0$ for $\phi_+=0$.
In this section, we calculate the profile of the inter-membrane distance
between two membranes which are quenched below the phase separation
temperature.

We first expand the potential $v(\ell; \phi_+=0)$ up to the fourth order terms
in $\delta$;
\begin{equation}
v(\ell; \phi_+=0) \approx v(\ell_0) + \frac{1}{2}V \delta^2 +
\frac{1}{4}U \delta ^4,
\end{equation}
where $V \equiv v''(\ell_0)\ell_0^2$  as before and
$U \equiv v^{(4)}(\ell_0)\ell_0^4 > 0$.
Both $V$ and $U$ are positive constants
because of the convexity of $v$ at its minimum.
Suppose that each of the membranes is in its high-temperature phase ($t>0$).
Then the fourth-order $\phi_i$ terms in the Ginzburg-Landau expansion
(\ref{glene2}) can be neglected since the second-order terms are
positive.
The resulting free energy with $\mu_+ = \mu_- = 0$ is
\begin{eqnarray}
F & = & \int \mbox{d}^2 \mbox{\boldmath $r$} \, \biggl[
\frac{1}{2} c \left[ (\nabla \phi_+)^2 + (\nabla \phi_-)^2 \right]
\nonumber  \\ & & +\frac{1}{2} t (\phi_+^2 + \phi_-^2) + \alpha
\phi_+ \delta \nonumber  \\ & & + \frac{1}{2} K (\nabla^2
\delta)^2  + \frac{1}{2} \Sigma (\nabla \delta)^2  + \frac{1}{2}V
\delta^2 + \frac{1}{4}U \delta^4 \biggr]  \nonumber  \\ & = &
\frac{1}{A}\sum_{\mbox{\boldmath $q$}} \, \biggl[ \frac{1}{2} (t
+c q^2) \big( \vert \tilde{\phi}_+(\mbox{\boldmath $q$}) \vert^2
+ \vert \tilde{\phi}_-(\mbox{\boldmath $q$}) \vert^2 \big)
\nonumber
\\ & & + \alpha \tilde{\phi}_+(\mbox{\boldmath $q$})
\tilde{\delta}(-\mbox{\boldmath $q$}) \nonumber  \\ & & +
\frac{1}{2} (V + \Sigma q^2 + K q^4) \vert
\tilde{\delta}(\mbox{\boldmath $q$}) \vert^2
 \biggr]
+ \int \mbox{d}^2 \mbox{\boldmath $r$} \, \frac{1}{4} U \delta^4.
\label{twominene}
\end{eqnarray}
We now minimize $F$ with respect to the concentrations
$\tilde{\phi}_+(\mbox{\boldmath $q$})$ and
$\tilde{\phi}_-(\mbox{\boldmath $q$})$ and obtain
\begin{equation}
\tilde{\phi}_+(\mbox{\boldmath $q$}) = - \frac{\alpha}{t+c q^2}
\tilde{\delta}(\mbox{\boldmath $q$}),~~~
\tilde{\phi}_-(\mbox{\boldmath $q$}) = 0.
\end{equation}
By inserting these equations into (\ref{twominene}) and expanding for
small $q$, the free energy can be written as
\begin{equation}
F = \int \mbox{d}^2 \mbox{\boldmath $r$} \, \biggl[ \frac{1}{2}
K_{\rm e} (\nabla^2 \delta)^2 +\frac{1}{2} \Sigma_{\rm e} (\nabla
\delta)^2 +\frac{1}{2} {V}_{\rm e} \delta^2 +\frac{1}{4} {U}_{\rm
e} \delta^4 \biggl], \label{memene}
\end{equation}
with
\begin{eqnarray}
{K}_{\rm e} \equiv K - \frac{\alpha^2 c^2}{t^3},~~~& &
{\Sigma}_{\rm e} \equiv \Sigma + \frac{\alpha^2 c}{t^2},
\nonumber  \\
{V}_{\rm e} \equiv V - \frac{\alpha^2}{t},~~~& &
{U}_{\rm e} \equiv U.
\end{eqnarray}
We see that for $t>0$ the coupling always
increases the mechanical tension $\Sigma_{\rm e} > \Sigma$, but reduces
the rigidity $K_{\rm e} < K$ and the potential strength $V_{\rm e} < V$.

Let us consider the strong coupling case when ${V}_{\rm e}<0$ but
still having ${K}_{\rm e}>0$, namely,
\begin{equation}
V < \frac{\alpha^2}{t}< K\left(\frac{t}{c}\right)^2.
\end{equation}
For $t>0$, although no phase separation occurs
in the absence of coupling ($\alpha=0$), it occurs
for non-zero $\alpha$.
The minimum free energy of the membranes is given by solving
the Euler-Lagrange equation obtained by minimizing (\ref{memene})
with respect to the inter-membrane distance $\delta$:
\begin{equation}
{K}_{\rm e} \nabla^4 \delta - {\Sigma}_{\rm e} \nabla^2 \delta
+{V}_{\rm e} \delta + {U}_{\rm e} \delta^3 = 0.
\label{euler}
\end{equation}
The two uniform (bulk) solutions of (\ref{euler}) are
\begin{equation}
\delta_0 = \pm \sqrt{-{V}_{\rm e}/{U}_{\rm e}}.
\label{uniform}
\end{equation}

We assume a one-dimensional profile $\delta(x)$ describing the
inter-membrane distance along the $x$-direction.
A typical profile determined by a numerical solution of the
Euler-Lagrange equation (\ref{euler}) using a relaxational method
is shown in Fig.\ \ref{profile}.
It is convenient to rescale the variables $\delta$ and $x$  as
$\zeta = {U}_{\rm e}^{1/3} \delta$ and
$u = {K}_{\rm e}^{-1/4} {U}_{\rm e}^{1/12} x$, respectively, 
yielding the following one-dimensional profile equation:
\begin{equation}
\frac{d^4 \zeta}{d u^4} -
\left(
\frac{{\Sigma}_{\rm e}}{{K}_{\rm e}^{1/2} {U}_{\rm e}^{1/6}}
\right)
\frac{d^2 \zeta}{d u^2}+
\left(
\frac{{V}_{\rm e}}{{U}_{\rm e}^{1/3}} \right)
\zeta + \zeta^3 = 0.
\end{equation}
Only two  independent combinations of the
four parameters $K_{\rm e}$ $\Sigma_{\rm e}$, $V_{\rm e}$
and $U_{\rm e}$ exist. In Fig.\ \ref{profile} they are set to be
${\Sigma}_{\rm e}/({K}_{\rm e}^{1/2} {U}_{\rm e}^{1/6})=0.1$ and
${V}_{\rm e}/{U}_{\rm e}^{1/3}=-1$, respectively.
The profile has a large slope at the interface $x=u=0$, but
relaxes to the bulk values $\pm \delta_0$ at $x = \pm \infty$
in a non-monotonic fashion with two symmetric overshoots,
having a height greater than $\delta_0$.
These overshoots are suppressed by increasing ${\Sigma}_{\rm e}$
or by increasing the coupling strength $\alpha$.
The maximum value of $\delta$ at the overshoot scales as
$\vert {V}_{\rm e} \vert^{1/2}$ as can be seen from (\ref{uniform}).
The overshoot of the profile is followed by a damped oscillation
which minimizes the curvature energy.
This behavior is similar to the nonlinear response of membranes to
local pinning sites \cite{BMMS,MS,MSK} or membranes adhering to
a geometrically structured substrate \cite{SA} and is a result
of the 4th order derivative in the profile equation.
The oscillatory decay has been also predicted for the membrane profile
between two inclusions such as proteins
\cite{DPS,DBPS,ABDPS,Netz}.

The configuration of the phase separated membranes corresponding
to the above inter-membrane distance $\delta$ is schematically
represented in Fig.\ \ref{overshoot}.
In the case of the adhesion of a single flexible membrane onto a
supported membrane, the supported membrane cannot have any shape
fluctuations.
Therefore, the inter-membrane distance profile calculated in this
section can be regarded as a distance of the flexible membrane
from the substrate with respect to its equilibrium distance $\ell_0$.

\section{Discussion}
\label{sec:discussion}

\subsection{Main Findings}

In this paper,  the interplay between adhesion and lateral phase
separation of multicomponent membranes is investigated. We
consider the ``bolaform-sticker'' adhesion where adhesive bridges
are formed by a single sticker having two sticky segments and
adhere directly onto the two membranes, as shown in Fig.\
\ref{figtype}(a). We proposed a phenomenological free energy
consisting of three parts: (i) the free energy  describing the
lateral phase separation of stickers on each membrane (see
(\ref{glene})); (ii) the deformation energy of the two membranes,
which is the sum of the bending energy, the surface tension, and
the potential energy (see (\ref{elene1})); and, (iii) the coupling
energy between the inter-membrane distance and the average
concentration of stickers on both membranes (see (\ref{symcoup})).
The difference of the chemical potentials between the two
membranes is also taken into account because the sticker
concentrations do not have to be the same.

We calculate the phase diagrams describing the bulk properties
for two particular choices of the chemical potentials, {\it i.e.},
$\mu_-=0~~ (\mu_1 = \mu_2)$ and $\mu_+=0~~ (\mu_1 = -\mu_2)$.
In the case of $\mu_-=0$, the critical temperature
increases depending on the coupling strength and
the potential strength (see (\ref{critical})).
Hence the lateral phase separation is {\it enhanced} due to the
adhesion.
This is one of the main consequences of our model.
When the phase separation takes place,
the inter-membrane distance is smaller for the domains rich in the
sticker molecules (``tight phase''), and  larger
for the domains poor in the stickers (``loose phase'').
In the case of $\mu_+=0$, our model exhibits a tricritical behavior.
The upward shift of the tricritical temperature also indicates the
{\it enhancement} of the lateral phase separation.

We find that the line tension for the lateral phase separation
increases because of the coupling effect as long as the mechanical
surface tension is non-zero.
We have also calculated the inter-membrane distance profile
between the two membranes which are quenched below their phase
separation temperature.
Because the membrane shape is governed by the
bending rigidity, the inter-membrane distance profile relaxes to
the bulk values in a non-monotonic way with two symmetric overshoots.

\subsection{Membrane Adhesion on Solid Surfaces and Supported Membranes}

So far, we have mainly discussed the adhesion of two membranes.
Our model also applies to the case where a single flexible membrane
with sticker molecules adheres to a flat substrate or a supported
membrane \cite{RFSS}.
Let us discuss these two cases separately.
For a flat substrate without any supported membrane on it, the
contributions from the second membrane (say $i=2$) can be dropped
from the model.
The coupling term (\ref{symcoup}) simply reduces to
$\alpha \phi_1 \delta$ because
the stickers are assumed to adhere directly to the substrate.
The second case is that of a supported membrane with sticker molecules.
Unlike the case of two fluctuating membranes discussed in 
Sec.\ \ref{profile}, the supported membrane does not have any
shape fluctuations. However,
even in such situations, there is an enhancement of the phase
separation due to the coupling effect and the upward shift of the
critical temperature is given by (\ref{critical}).
Another related situation is the case where a membrane is
composed of two different lipids and the membrane is put close to a
flat substrate.
If the two lipids feel different hydration force and prefer different
distances from the substrate, the phase separation between the two
components will be enhanced by the adhesion for the same reason
described in this paper.

\subsection{Relation to Other Models}

There exists an analogy between the phase behavior of our membrane system
with that of metamagnets (magnets which undergo fist-order phase
transitions in an increasing magnetic field) or $^3$He-$^4$He mixtures
described by the BEG (Blume-Emery-Griffith) spin-one
model \cite{BEG,KC}.
Moreover, the phase diagrams for $\mu_+=0$ are analogous to those
describing the phase separation of two-component mixtures in fluid
bilayers which also exhibits tricritical behavior \cite{MacS}
and other related amphiphilic systems \cite{LA,KK,HM,HMI,VA}.
However, in the former case of two-component bilayers
the concentration difference between the
two leaflets of the membrane is linearly coupled to the curvature of
the bilayer and the difference in the chemical potential is not taken
into account.

In our paper, we did not address the problem of the unbinding
transition. We rather assumed that the membranes are always bound
together, even in the absence of any sticker molecules. 
This assumption is partially motivated by the experimental study 
of Ref.\ \cite{BBS} where suspended membrane (part of the giant
vesicle) was claimed to be bound to the supporting membrane even
in the absence of sticker molecules.
In this case, the inter-membrane distance $\ell$ stays finite
and it is permissible to expand the free energy around the minimum.
Hence the phase separation consists of loosely and tightly
bound patches.
The interplay between unbinding transition and phase separation
of multi-component membranes has been considered in other
theoretical works \cite{Lipowsky96,Lipowsky97,WNL}.
The adhesion there is only brought about by sticker molecules, and
the phase separation is induced both by attractive interactions
and fluctuation-induced interactions between the stickers.
Although their model treats a different aspect of the more general
problem, the fluctuation effect yields similar consequences
compared to ours. We assumed that the $cis$-interaction in
(\ref{glene}) is attractive, and tracing over the inter-membrane
distance $\delta$ yields a term proportional to $\gamma \phi_+^2$.
Since this term does not depend on the sign of $\phi_+$, it has a
similar effect as fluctuations although our treatment is
restricted to the mean-field level.

It is worthwhile to comment here the difference between the
present study and that of Ref.\ \cite{BBS}. In their paper, it is
found that the adhesion between the membranes including homophilic
recognition molecules and repeller molecules is controlled by
lateral phase separation. The multiple competing states of
adhesion is attributed to the double-well inter-membrane
interaction potential generated by the competition of two forces;
attraction between homophilic molecules and the repulsion between
repeller molecules. By changing the repeller concentration, the
double-minimum potential causes the first-order transition between
a state with inter-membrane spacing set by the thickness of the
repeller molecules to a state with a spacing set by the bare
potential (van der Waals plus hydration interactions). In our
work, the effect of repeller molecules is not taken into account
and the minimum of the potential depends on the sticker
concentration through the coupling term (\ref{symcoup}). When the
stickers are phase separated and two different values of the
sticker concentration coexist, the inter-membrane potential has
double-minimum. However the physical origin of this double-minimum
potential is different from that in Ref.\ \cite{BBS} because it is
not due to the presence of repeller molecules.

\subsection{Other Types of Sticker Molecules}

As mentioned in the introduction, ``homophilic-sticker'' adhesion
occurs when the adhesive bridges are formed by two stickers of the
same type bound together by their two sticky segments (see Fig.\
\ref{figtype}(b)). Suppose $\psi_i$ ($i=1,2$) denotes the sticker
concentration on each membrane. Then the inter-membrane distance
depends on the product of each sticker concentration expressing
the probability to have two stickers -- one on each membrane -- at
the same position. Using (\ref{difference}) this coupling term can
be written in terms of $\phi_i$ as
\begin{equation}
\psi_1 \psi_2 = \phi_1 \phi_2 +\psi_c(\phi_1 +\phi_2)
+\psi_c^2.
\label{dcouple}
\end{equation}
An interesting remark can be made
for  homophilic-sticker adhesion. The resulting
phase separation within each membrane
leads to three different values for the inter-membrane distance.
The inter-membrane distance between domains rich in stickers
on both membranes
(rich-rich), as well as between rich-poor domains, and 
poor-poor domains can be different \cite{Lipowsky96,Lipowsky97}.
Notice that these three different inter-membrane distances correspond
to the tight, middle, and loose phases in our model.

A third case is that of ``lock-and-key'' adhesion due to the
formation of chemical bonds between lock-and-key types of
stickers, {\it e.g.,} ligands and receptors (see Fig.\
\ref{figtype}(c)). Suppose that both types of stickers are
distributed on the two membranes and $\psi_i$ now represents the
local concentration, say, of the lock molecules. First let us
assume that the membranes are saturated with sticker molecules (no
lipid). Then, $1-\psi_i(\mbox{\boldmath $r$})$ represents the
concentration of key molecules. Since domains rich in lock (key)
molecules on one membrane adhere with domains rich in key (lock)
molecules on the other membrane, the coupling tern in the free
energy $F_3$ will be a coupling between the inter-membrane
distance and
\begin{eqnarray}
& & \psi_1 (1-\psi_2)+(1-\psi_1)\psi_2  \nonumber  \\
& = & -2\phi_1 \phi_2 +(1-2\psi_c)(\phi_1+\phi_2)+2\psi_c(1-\psi_c).
\label{lkcouple}
\end{eqnarray}
This term is symmetric with respect to the exchange of two membranes.
Both in (\ref{dcouple}) and (\ref{lkcouple}), we see that the lowest order
term in the concentration (except the constant term) is proportional to
$\phi_+$.
If there is a linear coupling between the inter-membrane distance and
$\phi_+$ in these cases, we expect an upward shift in the transition
temperature and the phase separation will be enhanced as argued above.
Due to the presence of higher order terms in (\ref{dcouple}) and
(\ref{lkcouple}), however, the phase behavior 
will be more complex.

Let us now take into account the presence of lipids
in the lock-and-key adhesion.
If a single type of sticker is present on each membrane, namely,
lock molecules on membrane 1 and key molecules on membrane 2, we
can regard $\psi_1$ and $\psi_2$ as the concentrations of lock and key
molecules embedded in the lipid membrane, respectively.
Then $1-\psi_1$ and $1-\psi_2$ describe the concentration of the
second component (lipid) on each membrane, respectively.
In this case, the inter-membrane distance depends on $\psi_1 \psi_2$ as
in (\ref{dcouple}).
When both lock and key molecules are present on both membranes, one has
to start with a three-component mixture for each of the membranes.
The generic lattice model to study the behavior of ternary membranes
of monolayers is the BEG spin-one model \cite{BEG,KC,KGL}.
Here one has to include the coupling between the two membranes.
If we denote the concentration of lock and key molecules on each membrane
as $\psi^L_i$ and $\psi^K_i$ (and hence the concentration of the dilution
lipid
is $1 -\psi^L_i -\psi^K_i$), the inter-membrane distance now depends on
$\psi^L_1 \psi^K_2 + \psi^K_1 \psi^L_2$ which is similar to
(\ref{lkcouple}).
More detailed calculations for the homophilic stickers and lock-and-key
stickers and their influence
on membrane adhesion are left for future studies.

\acknowledgments

We have greatly benefited from the discussions and correspondence
with R.\ Netz, J.\ R\"adler, E.\ Sackmann, and T.\ Weikl.
SK would like to thank the Ministry of Education, Science and
Culture, Japan for providing financial support during his visit to
Israel.
Support from the exchange program between the Japan Society for
the Promotion of Science (JSPS) and the Israel Ministry of Science and
Technology is also gratefully acknowledged.
DA acknowledges partial support from the Israel Science Foundation
founded by the Israel Academy of Sciences and Humanities, Centers
of Excellence Program and the US-Israel Binational Science Foundation
(BSF) under grant number 98-00429.




\newpage
\begin{figure}
\caption{
Schematic representation of various types of adhesion between 
two membranes.
The two membranes are represented by black lines.
(a) Bolaform-sticker adhesion:
bridges consist of a single type of sticker molecules
which are anchored to one membrane (filled circle) and stick
to the other membrane by another sticky part of the molecules
(open circle).
(b) Homophilic-sticker adhesion: bridges consist of two
identical stickers which are
bound together by their respective sticky end segments.
(c) Lock-and-key adhesion: bridges consist of two different 
stickers forming a ligand-receptor type bond.
}
\label{figtype}
\end{figure}

\begin{figure}
\caption{
Two adhering membranes.
The reference $x$-$y$ plane is shown as a dashed line.
The height of two membranes measured from this plane is
denoted as $\ell_1$ and
$\ell_2$, respectively.
The sticker concentration on each membrane is denoted by
$\psi_1$ and
$\psi_2$, respectively.
}
\label{figmem}
\end{figure}

\begin{figure}
\caption{
Schematic phase diagram for a single membrane containing sticker
molecules.
The concentration of the sticker molecule is $\psi$.
The critical concentration and the critical temperature is denoted
by $\psi_c$ and $T_c$, respectively.
Within the coexistence curve the membrane separates
into sticker-rich and sticker-poor region (A+B coexistence).
The membrane is in a one-phase outside the coexistence curve.
}
\label{figphase}
\end{figure}

\begin{figure}
\caption{
Schematic representation of two adhering membranes undergoing
a lateral phase separation in the
case of the bolaform-sticker adhesion.
Coexistence between  tight (T) and loose (L) membrane domains is shown.
The inter-membrane distance is smaller than $\ell_0$
for the tight phase, whereas it larger than $\ell_0$
for the loose phase.
}
\label{figsym}
\end{figure}

\begin{figure}
\caption{
The phase diagram for the bolaform-sticker adhesion when $\mu_-=0$
as a function of (a) rescaled average composition
$\phi_+/\gamma^{1/2}$ and temperature $t/\gamma$, and (b) rescaled average
chemical potential $\mu_+/\gamma^{3/2}$ and temperature $t/\gamma$.
The continuous line is a first-order phase transition line, whereas the
dotted line is the limit of metastability of the asymmetric phase
with $\phi_- \neq 0$.
The metastable region of the asymmetric phase is indicated by a thick
line in (b).
Open circle ($\circ$) indicates a critical point.
Below the critical temperature, there is a coexistence region between
the loose (L) and the tight (T) phases as denoted by L+T.
The phase diagram is symmetric with respect to both
$\phi_+ \rightarrow -\phi_+$ and $\mu_+ \rightarrow -\mu_+$.
}
\label{figphasesym}
\end{figure}

\begin{figure}
\caption{
The phase diagram for the bolaform-sticker adhesion when $\mu_+=0$
as a function of (a) rescaled composition difference
$\phi_-/\gamma^{1/2}$ and temperature $t/\gamma$,
and (b) rescaled chemical potential difference $\mu_-/\gamma^{3/2}$ and
temperature $t/\gamma$.
The continuous line is a first-order phase transition line, whereas the
dashed line is a second-order one. The loose, tight and middle phases
are denoted as L, T, and M, respectively. The two tricritical points
are indicated by a filled circle ($\bullet$). Below them
there are two regions of coexistence of the middle phase ($\phi_+=0$)
and the tight phase ($\phi_+ \neq 0$) denoted as
M$_1$+T and T+M$_2$.
The phase diagram is symmetric with respect to both
$\phi_- \rightarrow -\phi_-$ and $\mu \rightarrow -\mu_-$.
Because $\mu_+=0$ there is a degeneracy between the tight (T)
and the loose (L) phases on the phase diagram.
}
\label{figphaseasym}
\end{figure}

\begin{figure}
\caption{
Schematic drawing of the inter-membrane distance
of two adhering
membranes with coexisting domains between:
(a) tight (T) and middle (M) phases, and;
(b)  loose (L) and middle (M) phases, as applies from Fig.\ 6
for bolaform-sticker adhesion.
The inter-membrane distance is exactly $\ell_0$ for the middle
phase, whereas it is smaller (larger) than $\ell_0$ for the tight
(loose) phase.
}
\label{figasym}
\end{figure}

\begin{figure}
\caption{
Inter-membrane distance profile at the domain boundary.
The variables are rescaled as
$\zeta = {U}_{\rm e}^{1/3} \delta$ and
$u = {K}_{\rm e}^{-1/4} {U}_{\rm e}^{1/12} x$.
The values of the two independent parameters are
chosen to be
${\Sigma}_{\rm e}/({K}_{\rm e}^{1/2}
{U}_{\rm e}^{1/6})=0.1$ and
${V}_{\rm e}/{U}_{\rm e}^{1/3}=-1$.
}
\label{profile}
\end{figure}

\begin{figure}
\caption{
Schematic drawing of the domain boundary
and inter-membrane distance of two adhering membranes.
Two symmetric overshoots in the interface region between the loose
and tight domains are shown, in agreement with the model and 
Fig.\ 8.
They result from the combined effect of membrane curvature and lateral
tension.
}
\label{overshoot}
\end{figure}


\newpage
\begin{figure}[tbh]
\epsfxsize=17cm
\centerline{\vbox{\epsffile{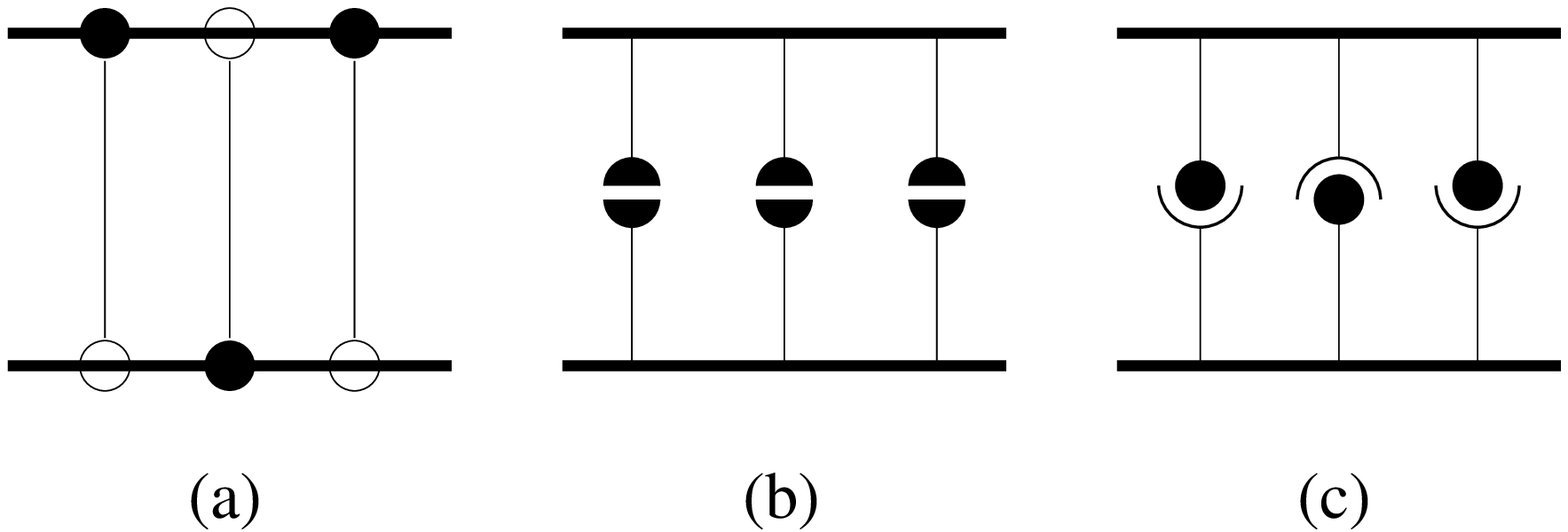}}}
\end{figure}
\vskip 4cm
 {\LARGE Fig.1  Komura and Andelman}

\newpage
\begin{figure}[tbh]
\epsfxsize=17cm
\centerline{\vbox{\epsffile{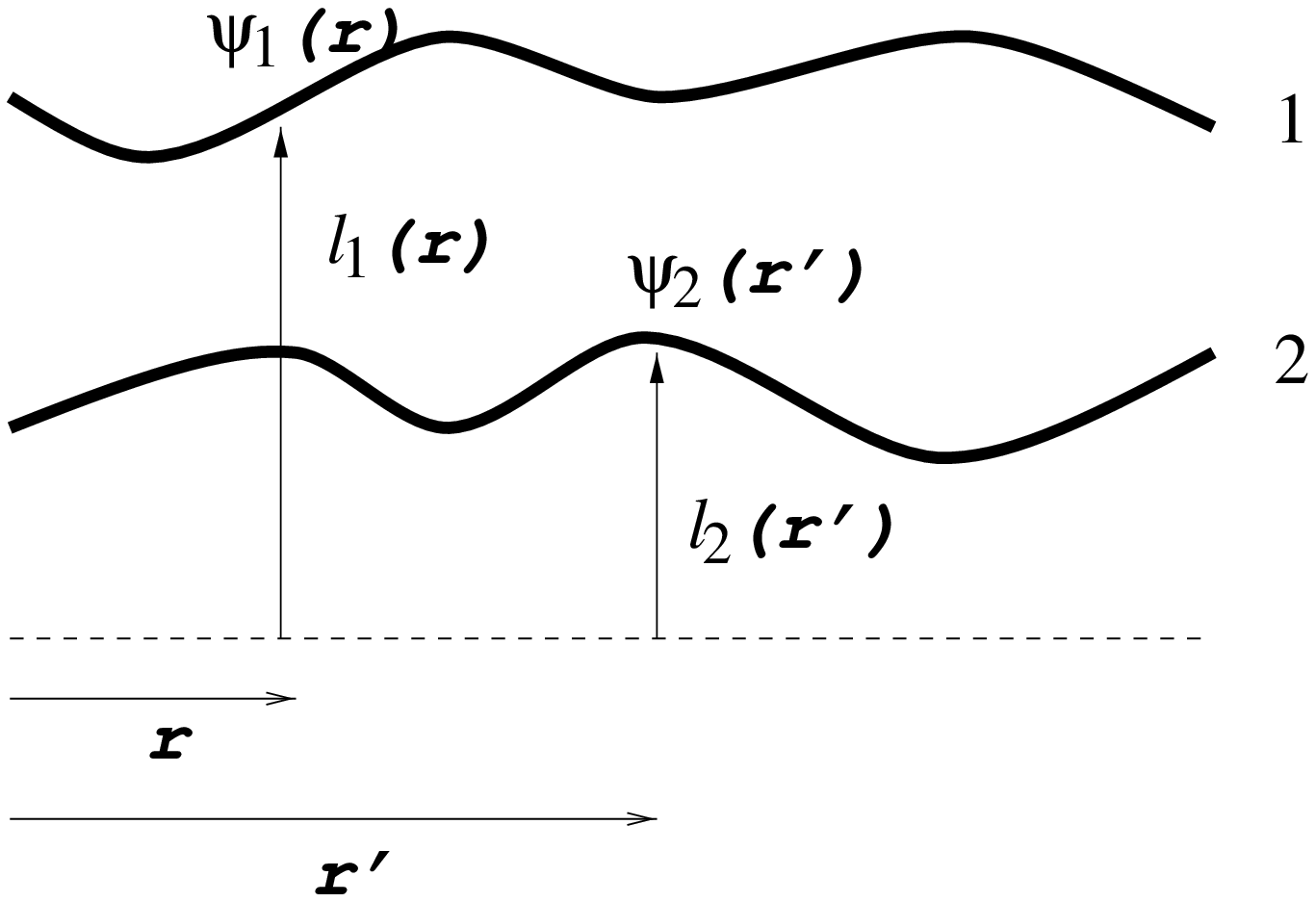}}}
\end{figure}
\vskip 4cm {\LARGE Fig.2  Komura and Andelman}

\newpage
\begin{figure}[tbh]
\epsfxsize=17cm
\centerline{\vbox{\epsffile{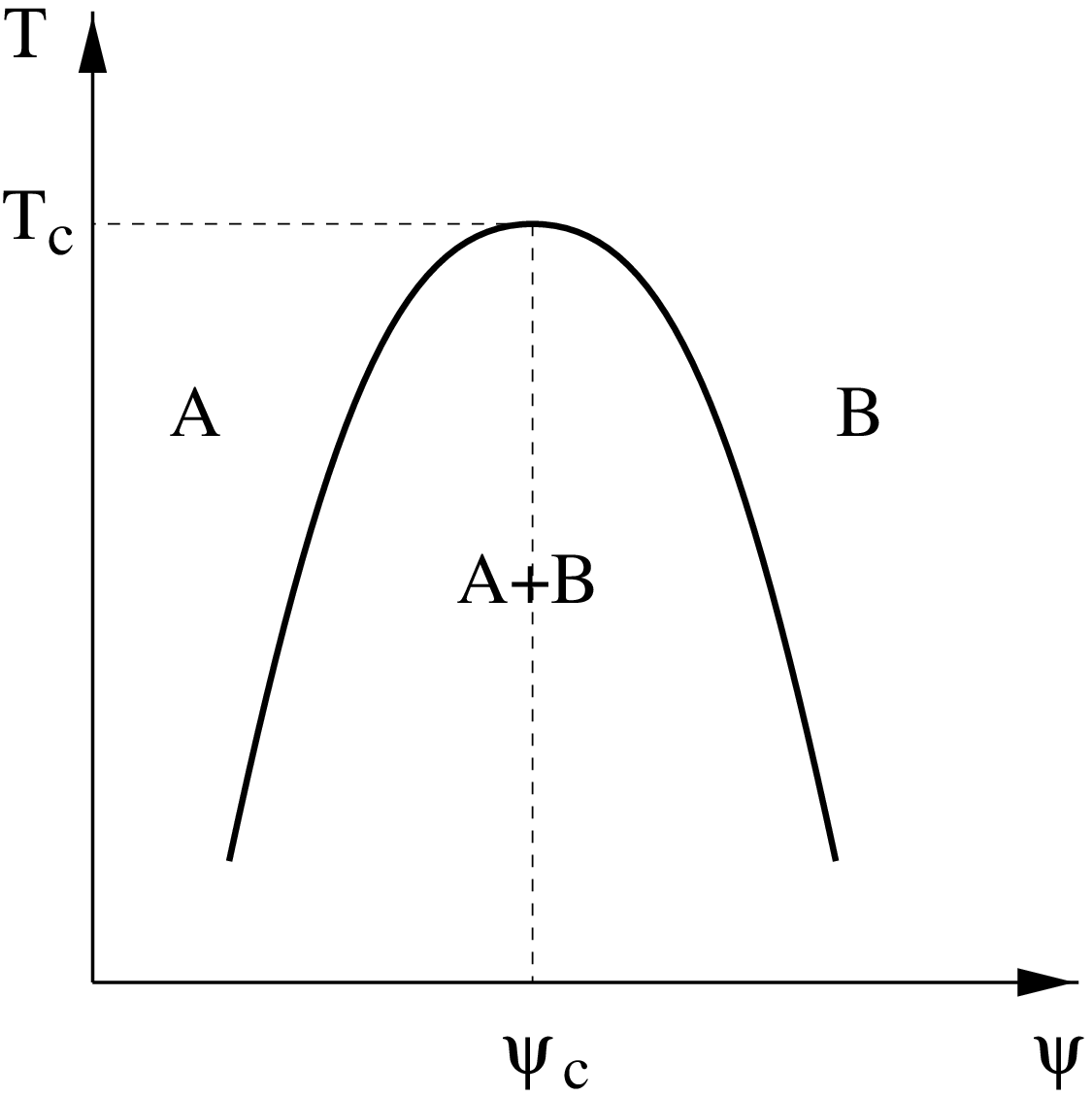}}}
\end{figure}
\vskip 4cm
{\LARGE Fig.3  Komura and Andelman}

\newpage
\begin{figure}[tbh]
\epsfxsize=17cm
\centerline{\vbox{\epsffile{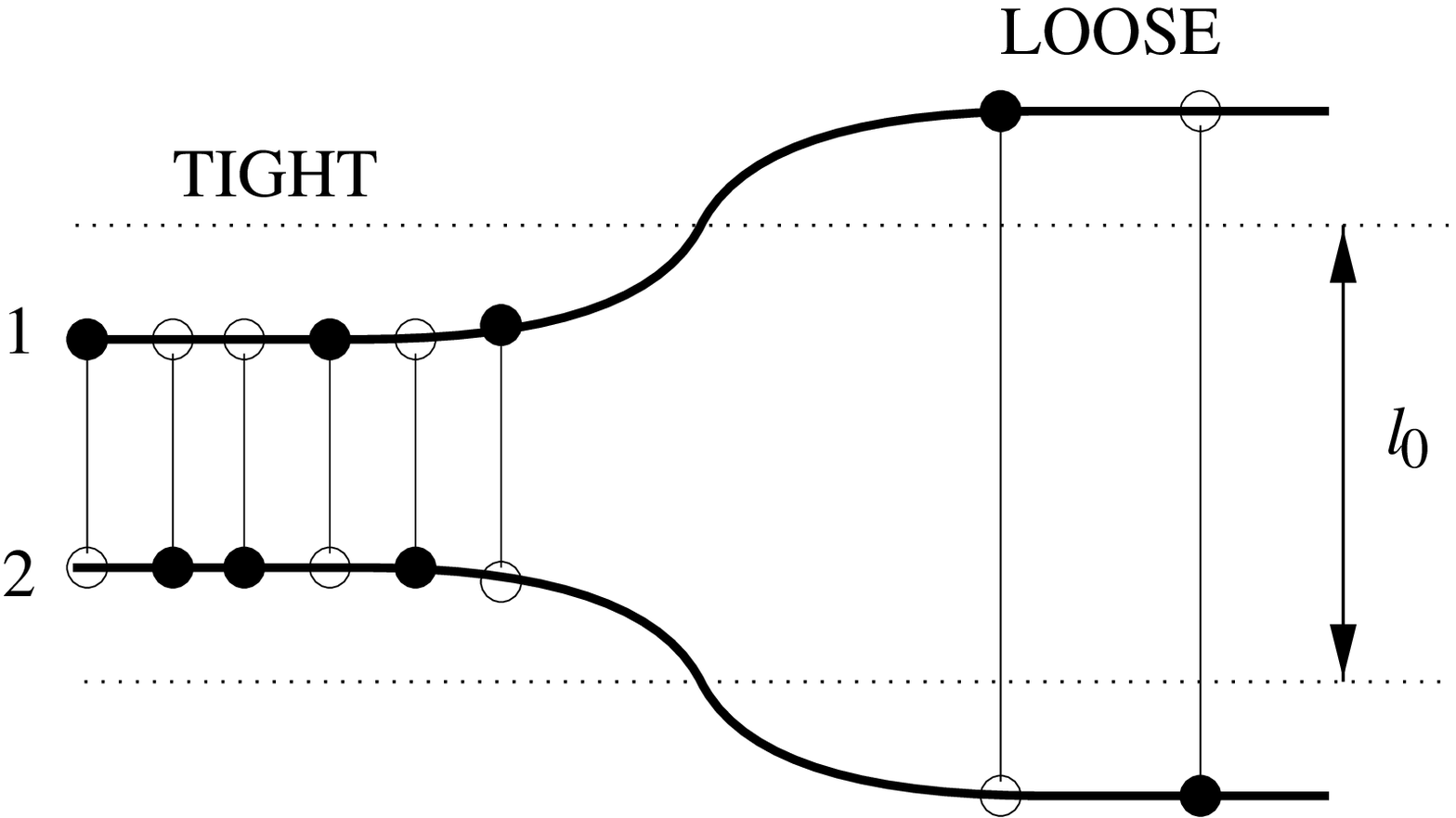}}}
\end{figure}
\vskip 4cm
{\LARGE Fig.4  Komura and Andelman}

\newpage
\begin{figure}[tbh]
\epsfxsize=17cm
\centerline{\vbox{\epsffile{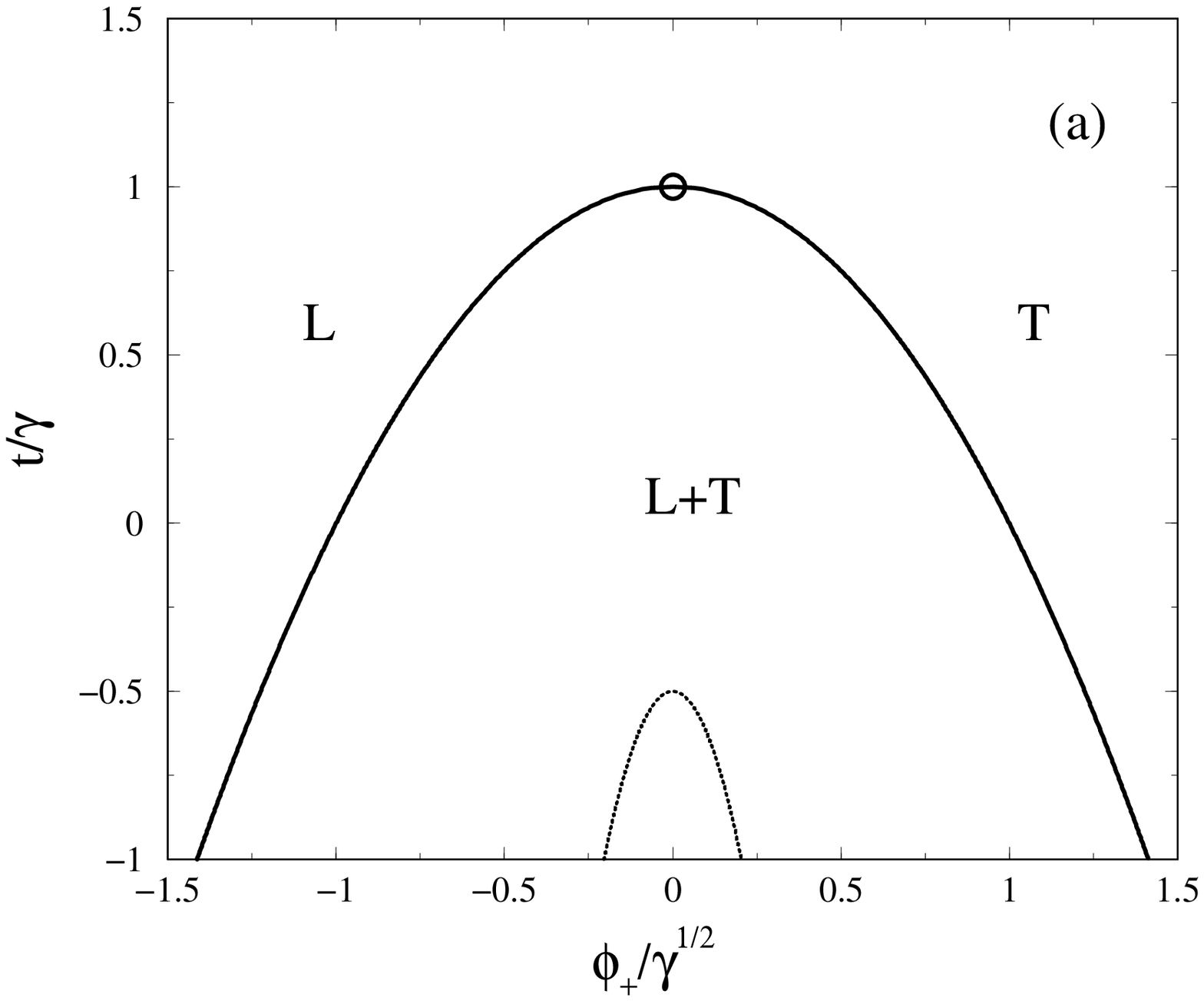}}}
\end{figure}
\vskip 4cm
{\LARGE Fig.5(a)  Komura and Andelman}

\newpage
\begin{figure}[tbh]
\epsfxsize=17cm
\centerline{\vbox{\epsffile{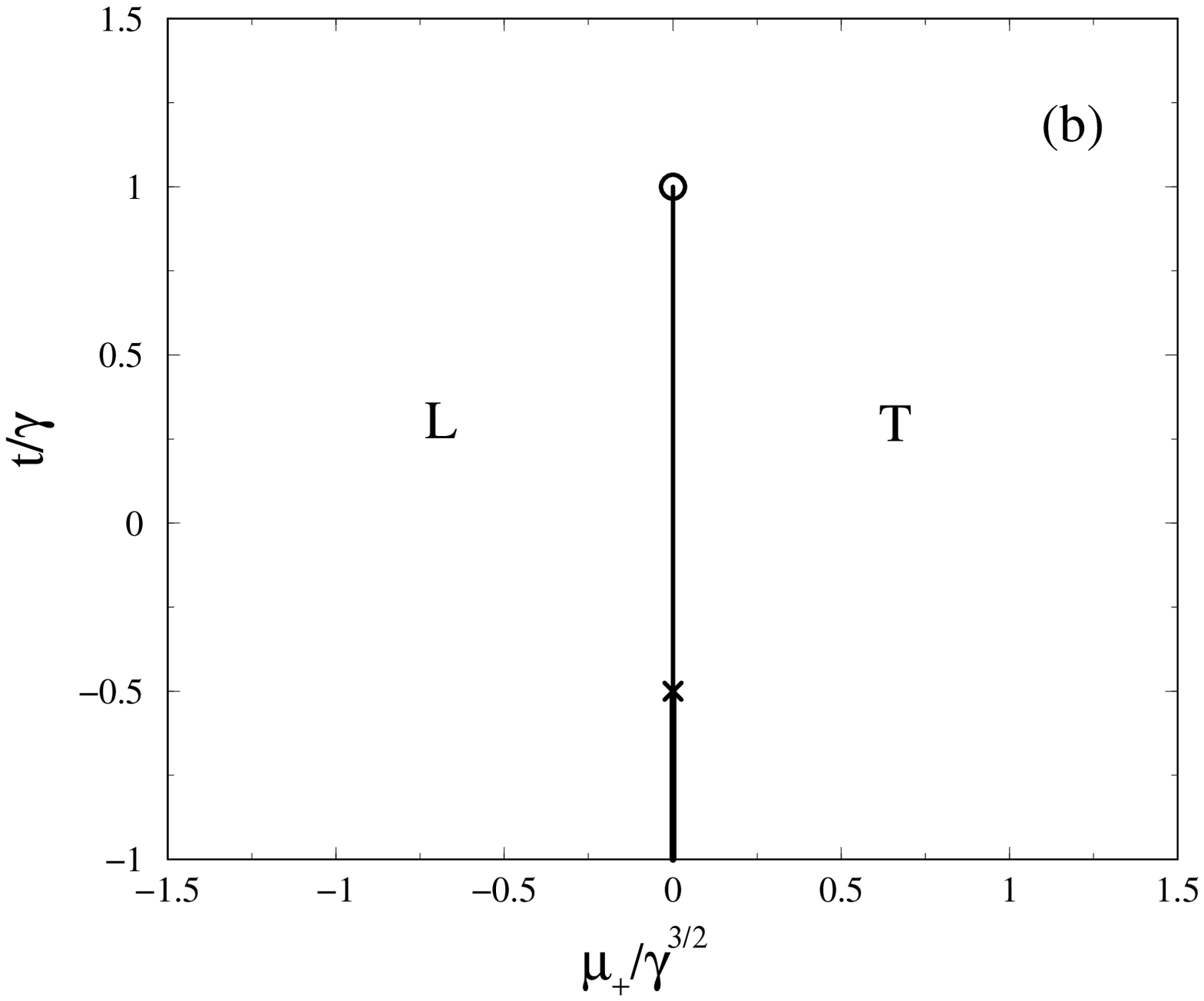}}}
\end{figure}
\vskip 4cm
{\LARGE Fig.5(b)  Komura and Andelman}

\newpage
\begin{figure}[tbh]
\epsfxsize=17cm
\centerline{\vbox{\epsffile{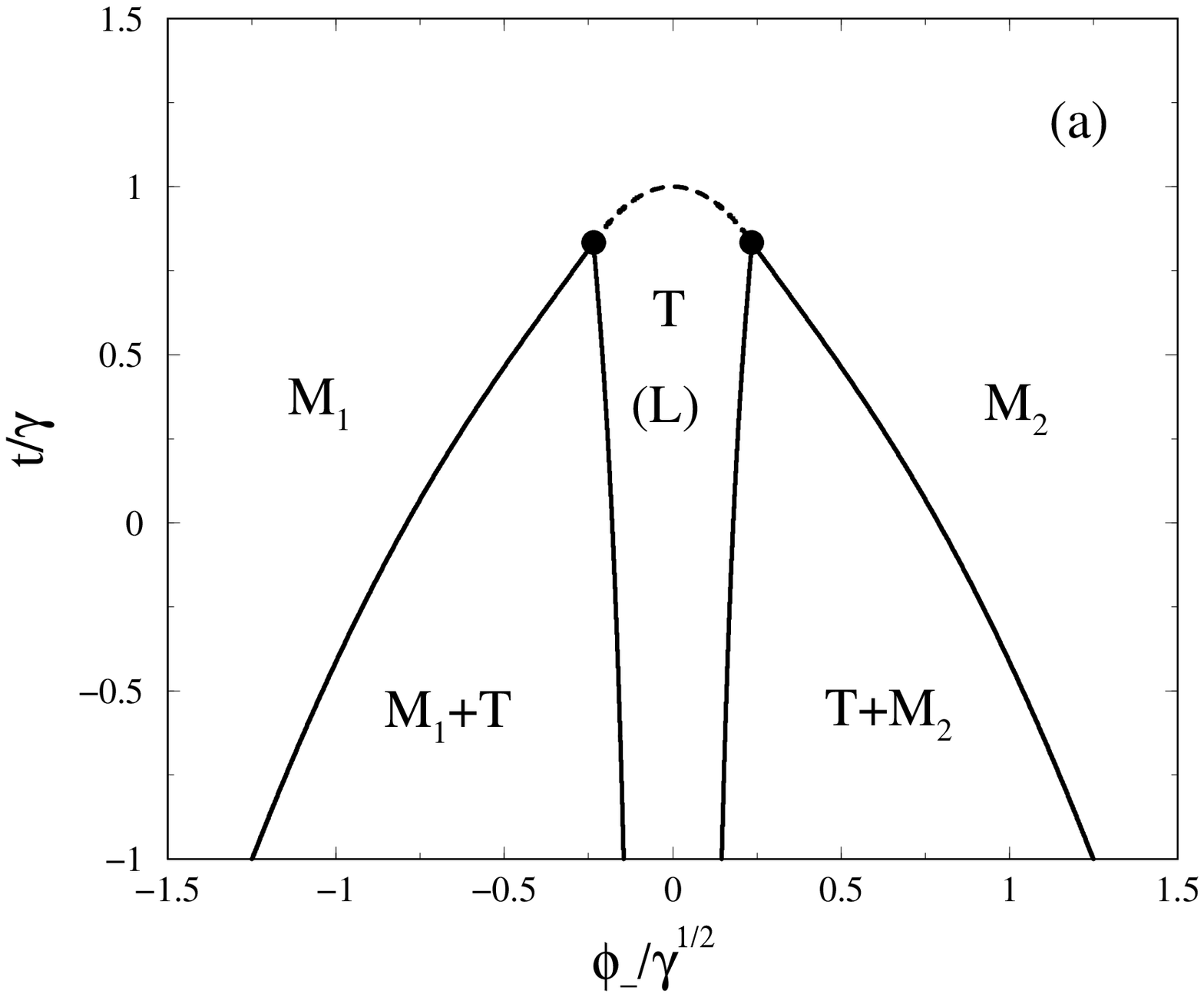}}}
\end{figure}
\vskip 4cm
{\LARGE Fig.6(a)  Komura and Andelman}

\newpage
\begin{figure}[tbh]
\epsfxsize=17cm
\centerline{\vbox{\epsffile{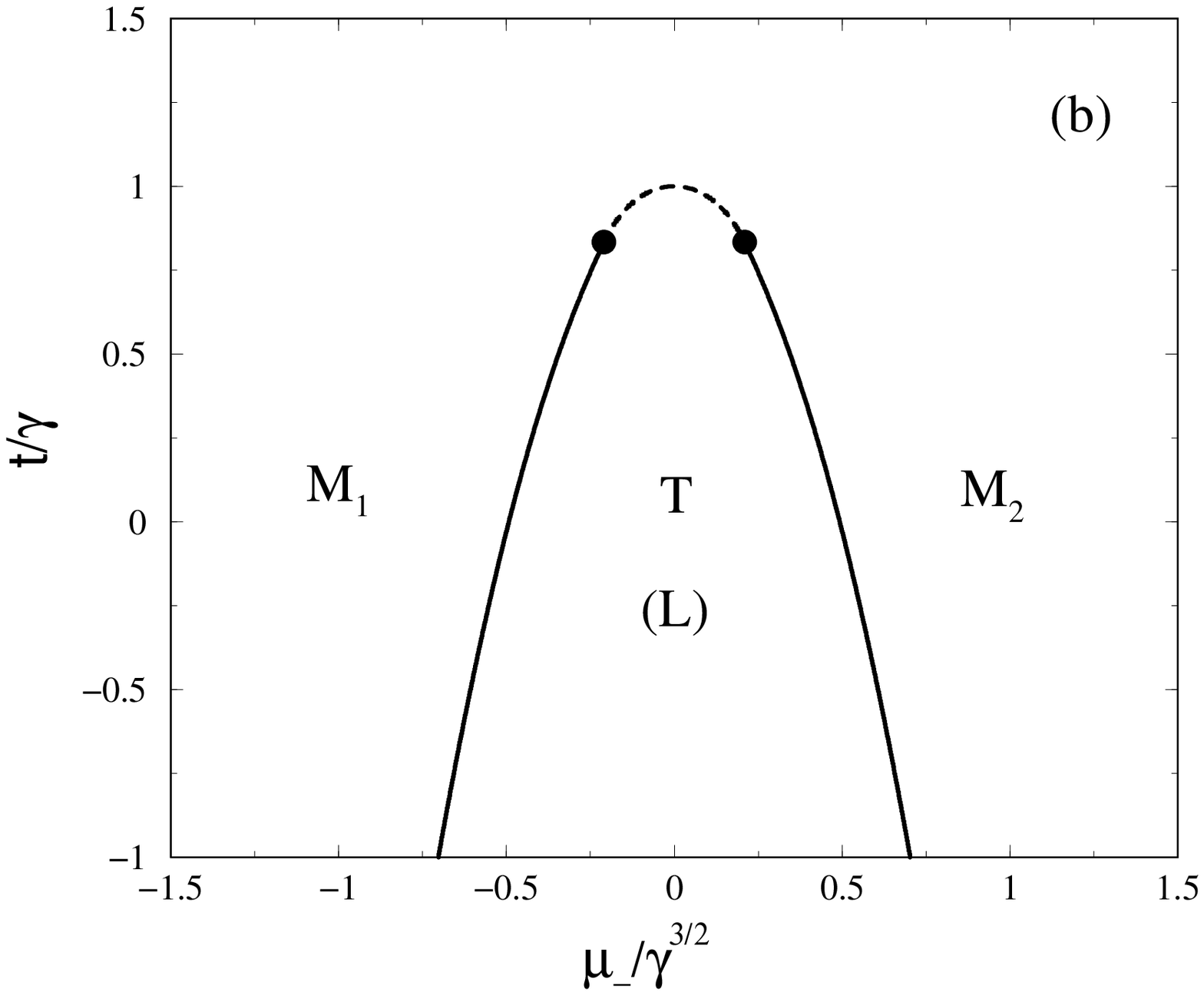}}}
\end{figure}
\vskip 4cm
{\LARGE Fig.6(b)  Komura and Andelman}

\newpage
\begin{figure}[tbh]
\epsfxsize=17cm
\centerline{\vbox{\epsffile{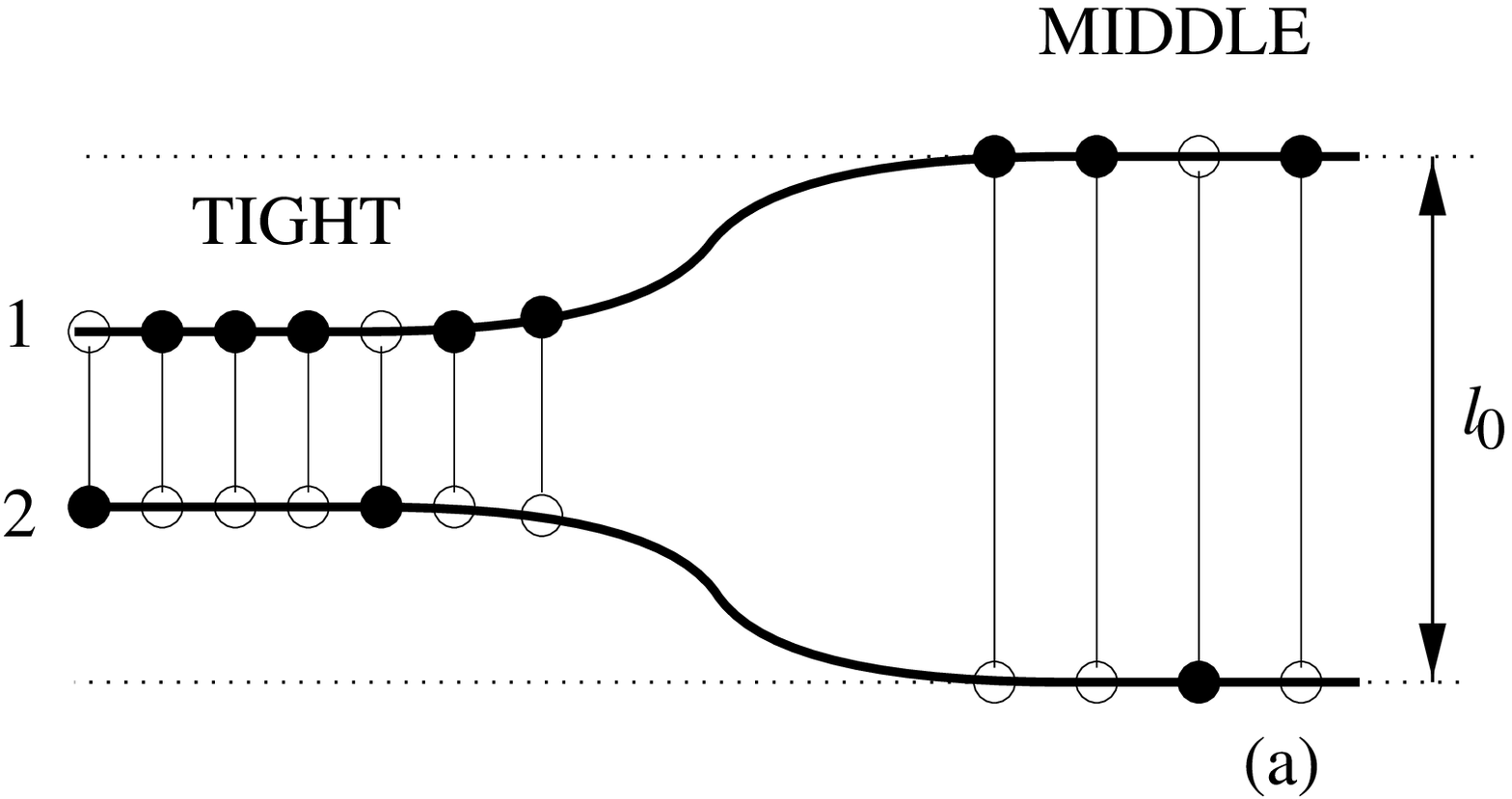}}}
\end{figure}
\vskip 4cm
{\LARGE Fig.7(a)  Komura and Andelman}

\newpage
\begin{figure}[tbh]
\epsfxsize=17cm
\centerline{\vbox{\epsffile{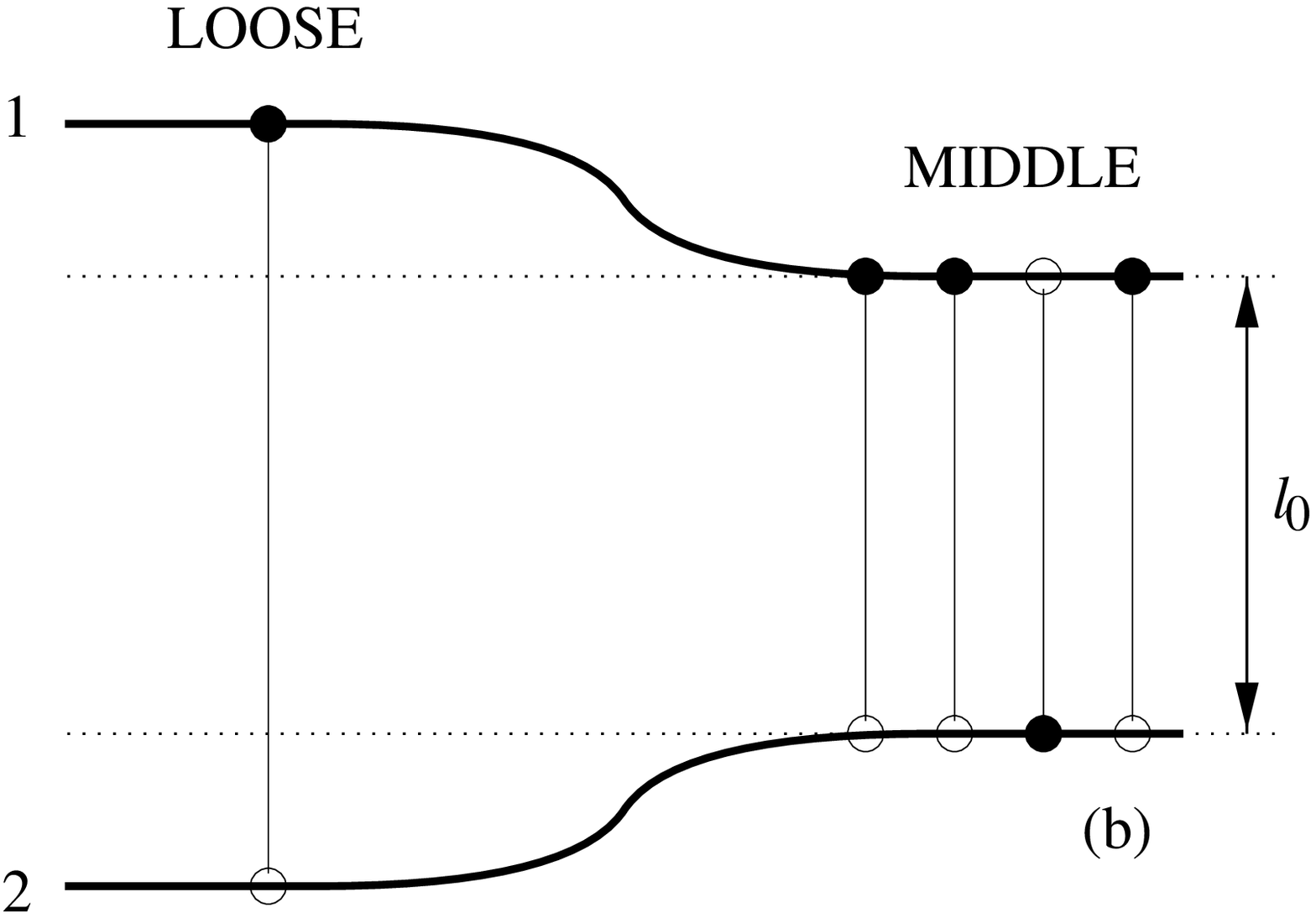}}}
\end{figure}
\vskip 4cm
{\LARGE Fig.7(b)  Komura and Andelman}

\newpage
\begin{figure}[tbh]
\epsfxsize=17cm
\centerline{\vbox{\epsffile{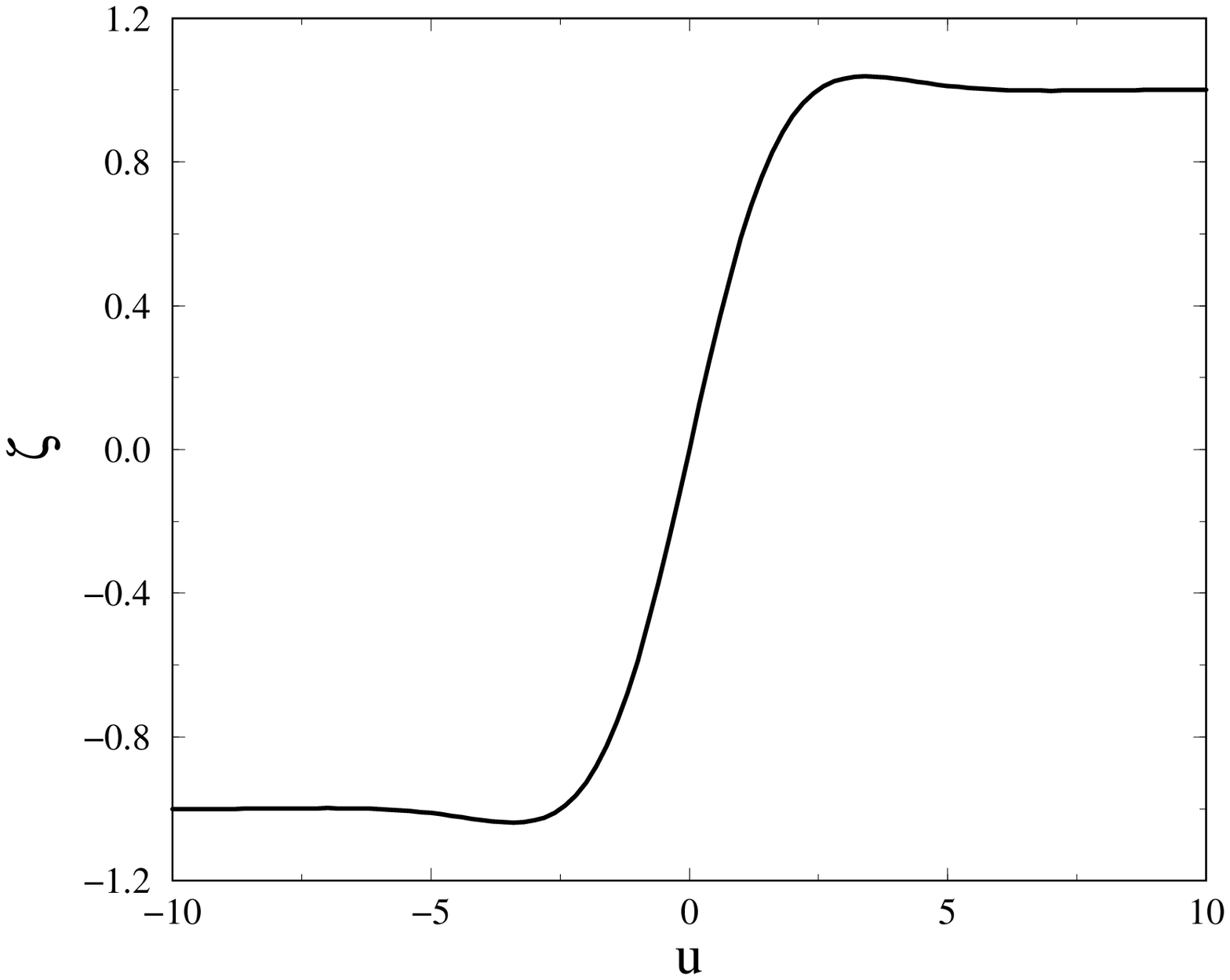}}}
\end{figure}
\vskip 4cm
{\LARGE Fig.8  Komura and Andelman}

\newpage
\begin{figure}[tbh]
\epsfxsize=17cm
\centerline{\vbox{\epsffile{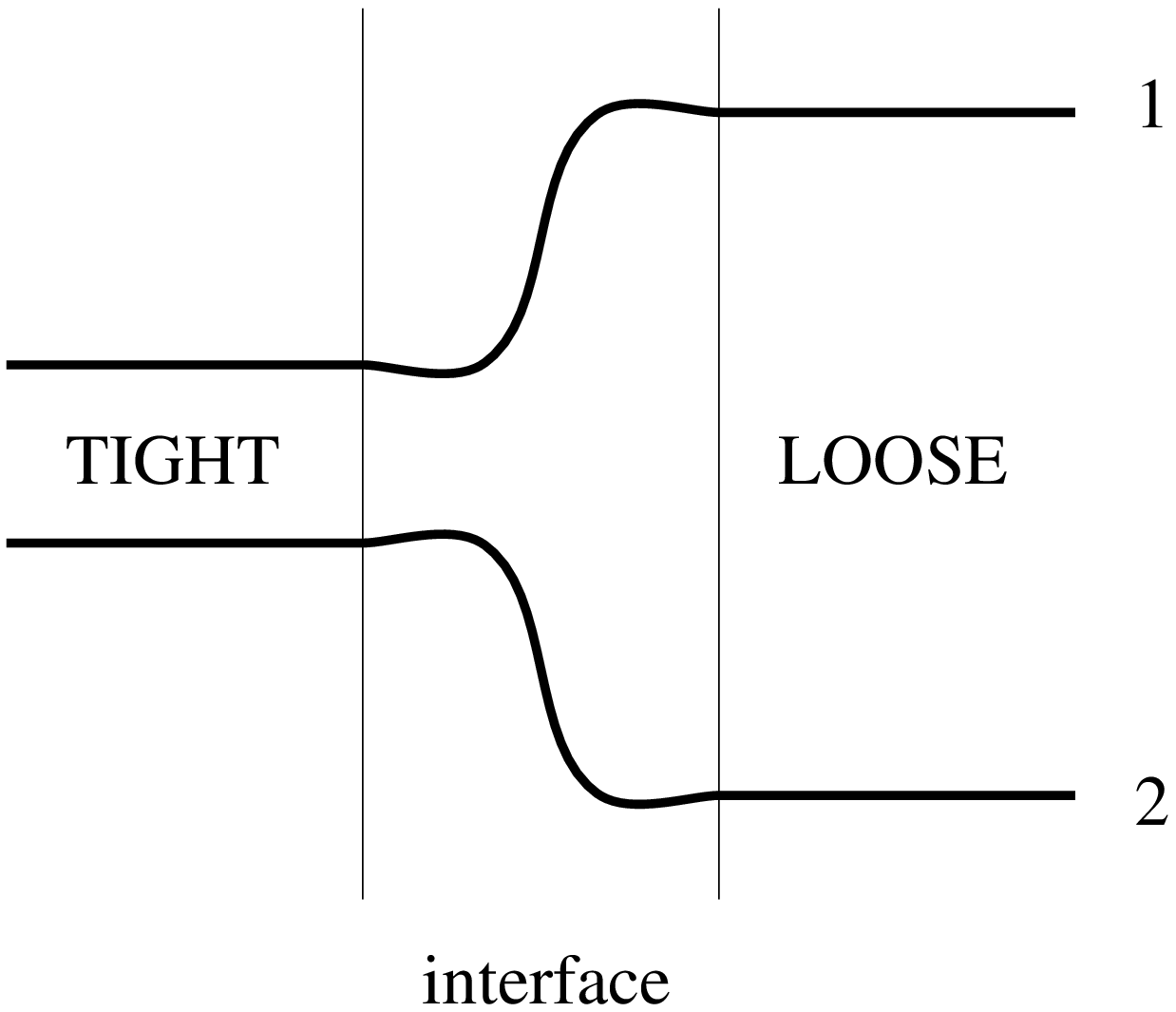}}}
\end{figure}
\vskip 4cm
{\LARGE Fig.9  Komura and Andelman}


\begin{references}
\bibitem{rigidity}
If the two membranes have different rigidity,
$\kappa_1$ and $\kappa_2$, the effective rigidity
is given by
$\kappa = \kappa_1 \kappa_2/(\kappa_1 + \kappa_2)$.
Similar relation holds for the surface tension.
See Ref.\ \cite{Lipowsky96}.
\bibitem{numbers}
When the minimum of the potential is determined only by
the balance between hydration and van der Waals interactions
(in the absence of any sticker molecules),
typical values are given by
$\ell_0 \approx 3 \times 10^{-9}$m,
$v(\ell_0) \approx -5 \times 10^{-6}$J/m$^{2}$, and
$V \approx 2 \times 10^{-4}$J/m$^{2}$.
See Ref.\ \cite{SA}.
\end{references}
\end{document}